\documentclass[aps,twocolumn,showpacs,amsmath,amssymb,pra,superscriptaddress,floatfix]{revtex4-1}

\usepackage{graphicx}
\usepackage{dcolumn}
\usepackage{bm}
\usepackage{amssymb}
\usepackage{multirow}
\usepackage{braket}
\usepackage{leftidx}

\begin{document}

\title{Energy dependent $\ell$-wave confinement-induced resonances}

\author{Benjamin He{\ss}}
    \email{bhess@physnet.uni-hamburg.de}
    \affiliation{Zentrum f\"ur Optische Quantentechnologien, Universit\"at Hamburg, Luruper Chaussee 149, 22761 Hamburg, Germany}

\author{Panagiotis Giannakeas}
    \email{pgiannak@physnet.uni-hamburg.de}
    \affiliation{Zentrum f\"ur Optische Quantentechnologien, Universit\"at Hamburg, Luruper Chaussee 149, 22761 Hamburg, Germany}

\author{Peter Schmelcher}
    \email{pschmelc@physnet.uni-hamburg.de}
    \affiliation{Zentrum f\"ur Optische Quantentechnologien, Universit\"at Hamburg, Luruper Chaussee 149, 22761 Hamburg, Germany}
    \affiliation{The Hamburg Centre for Ultrafast Imaging, Universit\"at Hamburg, Luruper Chaussee 149, 22761 Hamburg, Germany} 

\date{\today}


\begin{abstract}
The universal aspects of two-body collisions in the presence of a harmonic confinement are
investigated for both bosons and fermions. The main focus of this study are the
confinement-induced resonances (CIR) which are attributed to different angular momentum states
$\ell$ and we explicitly show that in alkaline collisions only four universal $\ell$-wave CIRs
emerge given that the interatomic potential is deep enough. Going beyond the single mode regime the energy dependence of $\ell$-wave CIRs is studied.
In particular we show that all the $\ell$-wave CIRs may emerge even when the underlying two-body potential
cannot support any bound state. We observe that the intricate dependence on the energy yields
resonant features where the colliding system within the confining potential experiences an effective
free-space scattering. Our analysis is done within the framework of  the generalized $K$-matrix theory
and the relevant analytical calculations are in very good agreement with the corresponding ab initio
numerical scattering simulations.
\end{abstract}

\maketitle

\section{Introduction}
\label{sec:introduction}
In low-dimensional ultracold gases the manipulation of two-body collisions preludes the
experimental realization of exotic many-body phases, such as the Tonks-Girardeau and the
super-Tonks-Girardeau gas phases \cite{kinoshita2004,paredes2004,haller2009}.
The concept of reduced dimensionality possesses a central role in these systems providing additional
means to control the corresponding collisional events apart from the usual toolkit of Fano-Feshbach
resonances \cite{chin2010,inouye1998,kohler2006}.
More specifically, it was shown that free-space non-resonant collisional events can be enhanced in
the presence of waveguide-like trapping potentials \cite{olshanii1998,bergeman2003}.
This yields a particular type of Fano-Feshbach resonance, the so-called confinement-induced
resonances (CIRs) which occur when the scattering length is of the order of the length of the
transversal confinement \cite{yurovsky2008, dunjko2011}.
Recent experimental advances allowed to explore the corresponding physics of CIRs in quasi-one- and
quasi-two-dimensional (quasi-1D and quasi-2D) waveguide geometries
\cite{haller2010,sala2013,frohlich2011,gunter2005,moritz2005} or in mixed dimensional scattering processes \cite{lamporesi2010}.
Complementing the experimental studies, substantial theoretical efforts exhibit a kaleidoscope of
confinement-induced processes, such as dual \cite{kim2006} and higher partial wave CIRs
\cite{granger2004, giannakeas2012}, multichannel
\cite{moore2004,saeidian2008,saeidian2012,melezhik2011} or anharmonic CIRs
\cite{peng2011,sala2012,peano2005} and CIR molecule formation \cite{melezhik2009} or dipolar CIRs \cite{giannakeas2013, sinha2007, hanna2012}.
Further studies on CIR effects focus on the impact of various confining geometries, such as quasi-2D
either harmonic \cite{petrov2001,idziaszek2006} or square well \cite{zhang2013}, and lattice potentials
\cite{zhang2013b, cui2010, fedichev2004}, or collisions in mixed dimensions \cite{nishida2010}.
Evidently, most of the theoretical efforts focus either on single partial wave collisions or on the low energy regime.
Therefore, extending the concept of mutually coupled higher partial wave CIRs to the regime of
strong energy dependence and investigating their universal properties will elucidate the underlying
collisional physics.\par
In this work, we consider two-body collisions of either bosonic or fermionic symmetry in the
presence of an axially symmetric and harmonic waveguide. The atoms are allowed to perform collisions
with higher partial waves where we take into account that the total colliding energy is well above
the threshold of the ground state of the transversal confinement. The theoretical conceptualization
of the corresponding resonant phenomena is based on a {\it fully} analytical and non-perturbative
framework of the $K$-matrix approach \cite{giannakeas2012, giannakeas2013} including appropriate 
interatomic interactions with e.g. a van der Waals tail. This approach provides a generalization of
the works of Granger {\it et al} \cite{granger2004} and Kim {\it et al} \cite{kim2005}
incorporating however {\it all} the higher partial waves and contributions from {\it all} the closed
channels. Furthermore, going beyond the previous studies we derive the connection of the {\it
physical} $K$-matrix with all the relevant scattering observables obtaining thus the {\it full }
scattering wave function. Therefore, the $K$-matrix approach can be applied equivalently to other
systems, e.g. distinguishable or identical particles for various confining geometries or
multichannel collisions beyond the single mode regime. Investigating the univeral properties of the
$\ell$-wave CIRs for alkali atoms we observe that only four are present, namely $s$- ($p$-) and
$d$- ($f$-)wave for bosons (fermions) due to the interplay between confinement and
a sufficiently deep van der Waals potential.
In addition, we study the energy dependence of the corresponding CIRs and observe that all the
CIRs can occur even when the two-body potential cannot sustain a weakly bound or quasi-bound state.
Moreover, for the energy dependent $d$- and $f$-wave CIRs we observe the bosonic and fermionic
counterpart of the dual CIR \cite{kim2006}, i.e. total transparency ($T=1$), which occurs due to the
destructive interference of $s-d$ and $p-f$ wave scattering, respectively. Finally, ab initio
numerical simulations corroborate all the
relevant analytical calculations within the $K$-matrix framework.\par
In detail, this paper is organized as follows. In Sec. \ref{sec:model} we introduce the waveguide
Hamiltonian under consideration, define the interatomic potential and discuss the relevant theoretical
methods, namely the $K$-matrix theory for quasi-1D waveguide geometries. The connection between the
physical $K$-matrix and the physical observables is the content of Sec. \ref{sec:observables},
while Sec. \ref{sec:results} is devoted to the discussion of the universal properties of interatomic potential
exhibiting a van der Waals tail and how
this affects the CIRs in neutral diatomic collisions. A summary and conclusions are given in Sec. \ref{sec:conclusions}.

\section{Waveguide Hamiltonian and $K$-Matrix Approach}
\label{sec:model}
We consider the collision of two identical particles within a harmonic waveguide.
Due to the harmonicity of the confinement we can separate the Hamiltonian into a part describing
the center of mass and relative motion, respectively. The Hamiltonian of the relative motion
relevant to the collisional process reads
\begin{equation}
H=\frac{-\hbar^2}{2\mu}\Delta+\frac{1}{2}\mu\omega_{\perp}^2\rho^2+V_{LJ}(\mathbf{r}),
\label{eq:Hamiltonian}
\end{equation}
where $r=\sqrt{z^2+\rho^2}$ is the interparticle distance, with $z$ and $\rho$ describing the
longitudinal and transversal degrees of freedom, respectively.
$\mu$ denotes the reduced mass of the colliding pair and $\omega_\perp$ is the confinement frequency.
Accordingly, the harmonic oscillator length scale is given by $a_{\perp}=\sqrt{\hbar / \mu \omega_{\perp}}$.
The term $V_{LJ}(r)=\frac{C_{10}}{r^{10}}-\frac{C_{6}}{r^{6}}$ is the Lennard-Jones 6-10 potential
indicating the short-ranged two-body interatomic interactions.
$C_{6}$ is the dispersion coefficient and it defines the van der Waals length scale via the relation
$\beta_{6}=(2\mu C_{6}/\hbar^2)^{1/4}$.
We set $C_{10}$ as a parameter in order to tune the corresponding scattering lengths induced by the
short-range potential. Among others, the particular choice $V_{LJ}(r)$ is motivated by the existence
of analytical solution of the free-space collisional problem, where the corresponding
phase shifts are derived by means of a generalized effective range theory \cite{gao2009}.\par
Hereafter, we assume that the length scales in our Hamiltonian are well separated, namely  $\beta_{6}\ll a_{\perp}$.
Due to this separation, our configuration space has three distinct regions.\par
(i) In the asymptotic region $r\to \infty$ the interatomic potential $V_{LJ}$ is negligible.
Therefore, the wave function can be written as a linear combination of product states of plane waves
in $z$-direction and a 2D harmonic oscillator state in the transversal $\rho$-direction. The
relevant scattering information is then encapsulated in the $K^{1D}$ matrix.
In addition, the total colliding energy $E$ distributes over these two degrees of freedom according
to $E=\frac{(a_{\perp}k)^2}{2}=2n+1+\frac{(a_{\perp}q_{n})^2}{2}$, where $n$ refers to the
oscillator modes, $q_{n}$ denotes the momentum in the unconfined $z$-direction and $k$ indicates the
wave vector of the total energy.\par
(ii) Approaching the origin from the asymptotic region we pass through an intermediate regime ($\beta_6\ll r \ll a_\perp$), where
both the confining- and the interatomic- potential are negligible.
Hence, the wave function in the corresponding region is simply the solution of the free particle
Hamiltonian.\par
(iii) Finally, in the inner region ($r\sim \beta_6$), $V_{LJ}$ becomes the by far dominant contribution
and thus the two particles experience a free-space collision with total energy
$E=\frac{(a_{\perp}k)^2}{2}$ , where the corresponding wave function can be written as a linear
combination of the free-space solutions.
The impact of the $V_{LJ}$ potential is then encompassed in the $K^{3D}$ matrix.\par
The crucial assumption of length scale separation permits us to map the solutions from the inner region, where spherical
symmetry is present and states are thus characterized by angular momentum eigenstates $\ket{l}$, to
the states $\ket{n}$ in the asymptotic
regime, where $n$ denotes the $n$-\emph{th} oscillator mode.
We remark that in the inner and asymptotic region
the azimuthal subgroup of the spherical symmetry group remains a symmetry of the Hamiltonian and therefore no
mixing of different azimuthal states $m$ occurs.
Hence, we set $m=0$ and omit it in the labeling of the states.
The map accomplishing this local frame transformation from the spherical to the cylindrical
solutions was used before \cite{harmin1982,*harmin1982prl,*harmin1985electric,greene1987} and is
generally given by
\begin{equation}
U_{\ell n}=\frac{\sqrt{2}(-1)^{d_0}}{a_{\perp}}\sqrt{\frac{2\ell+1}{k
q_{n}}}P_{\ell}\Bigl(\frac{q_{n}}{k}\Bigr),
\label{eq:local_frame_transformation}
\end{equation}
where $d_{0}$ is either given by $\ell/2$ in the case of even partial waves, or, respectively by $(\ell+1)/2$ in the
case of odd partial waves and $P_{\ell}(\cdot)$ indicates the Legendre polynomial of $\ell$-th degree.
As mentioned above, Eq. (\ref{eq:local_frame_transformation}) interrelates the wave functions of
the asymptotic and the inner region.
Consequently, the corresponding $K$ matrices are connected according to the relation $K^{1D}=U K^{3D} U^{T}$
\cite{granger2004,giannakeas2012,giannakeas2013,zhang2013}, where the
$K^{3D}$-Matrix is a diagonal matrix in the $\ell$-wave representation with entries given by the tangent of the phase
shifts.\par
Allowing $K^{1D}$ and $K^{3D}$ to be fully energy dependent, in the following we consider that the total colliding
energy is well above the threshold of the transversal ground state and below the threshold to the first excited
transversal mode.
This implies that we assume only one open channel, namely the ground state of the harmonic oscillator, while all the excited states remain closed.
However the part of the wave function which refers to the closed channels possess exponential
divergences, and thus, results in an unphysical scattering process.
To make the excited modes become evanescent, i.e. impose the correct
physical boundary conditions the multichannel quantum defect theory is used, which was first utilized in
the context of CIRs in
\cite{granger2004} and is in more detail generally discussed in \cite{aymar1996multichannel}, leaving us with a physical
$K$-Matrix, given by
\begin{equation}
  K_{oo}^{1D,phys}=K_{oo}^{1D}+i K_{oc}^{1D}(1-iK_{cc}^{1D})^{-1}K_{co}^{1D},
  \label{eq:k_oo_1d_phys}
\end{equation}
where $K_{oo}^{1D}$ indicates the open-open channel transitions, $K_{oc}^{1D}$ and $K_{co}^{1D}$
refer to the K-matrices responsible for open-closed and
closed-open channel transitions, respectively, while $K_{cc}^{1D}$ denote the transitions between
closed channels.
According to Eq. \eqref{eq:k_oo_1d_phys} the resonant processes are manifested as poles of the physical $K^{1D}$-matrix.
Therefore, the roots of $\operatorname{det}(\openone-iK_{cc}^{1D})$ correspond to the positions
of closed channel bound states lying in the continuum of the open channel. This means that the
corresponding resonant structure fulfills a Fano-Feshbach scenario.\par
Before addressing the question of how to relate the physical $K$-Matrix from Eq. \eqref{eq:k_oo_1d_phys} to the physical observables, let us for convenience
introduce (c.f. App. \ref{app:derivation_umatrix}) the trace over the closed channels $\mathfrak{U}_{\ell\ell^\prime}$ of
$U_{\ell n}U_{\ell^\prime n}$, for arbitrary angular momenta $\ell$ and
$\ell^{\prime}$, by
\begin{widetext}
\begin{gather} 
  \mathfrak{U}_{\ell\ell^\prime}(\epsilon)=(-1)^{\frac{\ell+\ell^\prime}{2}+\sigma}\sqrt{(2\ell+1)(2\ell^\prime+1)}\sum_{\nu=|\ell-\ell^\prime|}^{\ell+\ell^\prime}\sum_{p=0}^{\nu}\frac{\Gamma(\ell,\ell^\prime,\nu,p)}{(\epsilon+\frac{1}{2})^{\frac{p+1}{2}}}\zeta(-\frac{p-1}{2},n_o-\epsilon),\quad\mbox{with}\\[3mm]
\Gamma(\ell,\ell^\prime,\nu,p)=i^{p-1}2^{\nu-1}\begin{pmatrix}\ell&\ell^\prime&\nu\\0&0&0\end{pmatrix}(2\nu+1)\begin{pmatrix}\nu\\p\end{pmatrix}\begin{pmatrix}\frac{\nu+p+1}{2}\\\nu\end{pmatrix}\nonumber
  \label{eq:umatrix}
\end{gather}
\end{widetext}
where $n_{o}$ denotes the total number of open channels, which in the following is equal to 1. $\zeta(s,q)$ denotes the Hurwitz
Zeta function, in the relation $\Gamma(\cdot)$ the terms which depend on $\ell,\ell^{\prime}$ and $\lambda$ denote the Wigner
$3J$-symbols, while the last two expressions denote the
binomial coefficients. $\epsilon$ is the dimensionless,
channel normalized, total colliding energy, given by the relation
$E=2\omega_{\perp}(\epsilon+\frac{1}{2})$. This scale is chosen in such a way, that for
$0\le\epsilon<1$, the colliding energy varies between the thresholds of the ground- and the first
excited transversal mode, respectively. It is clear from the definition
of the trace, that $\mathfrak{U}_{\ell\ell^\prime}$ describes the coupling of a partial wave
$\ell$ to a $\ell^\prime$-wave after undergoing
the virtual transitions in the closed channels $n$ which are encapsulated in the Hurwitz Zeta  
functions.\par
Note that the elements of $\mathfrak{U}$ can be used for different interatomic
potentials, as long as harmonic confinement is considered and the separation of length scales is
fulfilled. For example a minor generalization of them yields a convenient
representation for the couplings to the closed channels with possible applications in the case of dipolar collisions.\par
The actual derivation of the physical $K$-Matrix essentially reduces to the inversion of the part
containing the $cc$-channel contributions, i.e. $(\openone - iK^{1D}_{cc})^{-1}$.
Following the derivation in \cite{giannakeas2012}, we obtain a physical $K$-Matrix in the open channels, which
reads
\begin{align}
K_{oo}^{1D,phys}=\frac{1}{\det(\openone-iK^{3D}\mathfrak{U})}\times\Bigl(\Delta_\ell
U_{\ell0}^2+\Delta_{\ell^{\prime}}U_{\ell^{\prime}0}^2-\nonumber\\
-i\Delta_\ell\Delta_{\ell^{\prime}}(\mathfrak{U}_{\ell^{\prime}\ell^{\prime}}U_{\ell
0}^2+\mathfrak{U}_{\ell\ell}U_{\ell^{\prime}0}^2-2\mathfrak{U}_{\ell\ell^{\prime}}U_{\ell
0}U_{\ell^{\prime} 0})\Bigr),
\label{eq:k_oo_double}
\end{align}
where $K^{3D}\mathfrak{U}$ denotes the matrix product
$\sum_{\lambda}K^{3D}_{\ell\lambda}\mathfrak{U}_{\lambda\ell^{\prime}}$, while $\Delta_\ell = \tan\delta_{\ell}$, whereas in turn $\delta_{\ell}$ refers to the phase shift of the $\ell$-th
partial wave gathered in the interatomic collision via $V_{LJ}$.
We remark, that Eq. \eqref{eq:k_oo_double} holds for arbitrary
partial waves $\ell$ and $\ell^{\prime}$, under the assumption that either both are even or odd and
at all energies within the continuum between the ground- and first excited channel, where the main
difference to the $K$-Matrix derived in \cite{giannakeas2012} is, that the $K$-Matrix there
only holds as long as $a_{\perp}q_{0}\ll 1$. In particular we also note, that in the case
of $\Delta_{\ell^{\prime}}=0$, namely the single partial wave approximation, the formula reduces to
\begin{equation}
K_{oo}^{1D,phys}=\frac{\Delta_\ell U_{\ell 0}^2}{1-i\Delta_\ell\mathfrak{U}_{\ell\ell}},
\label{eq:k_oo_single}
\end{equation}
which resembles for $\ell=0$ and $\ell=1$ the results found in \cite{olshanii1998} and
\cite{granger2004}, respectively.
\section{$K$-Matrix and Observables}
\label{sec:observables}
In order to relate the physical $K$-Matrix to the relevant physical observables, different approaches
were chosen in the past. One of the most common techniques to achieve this goal was to introduce an
effective quasi 1D Hamiltonian to which the original Hamiltonian was mapped, where the short-range
interaction is modeled by a bare delta function multiplied by a factor which is expressed in terms
of the physical $K$-matrix \cite{olshanii1998,giannakeas2012}. This technique was subsequently
extended by the Bose-Fermi mapping to also include spin-polarized fermionic systems 
\cite{granger2004}. This procedure of mapping
to an auxiliary Hamiltonian becomes cumbersome in the multi mode-regime, resulting in
a family of effective Hamiltonians which are not uniquely defined.\par
To avoid this difficulty we present an alternative method which by construction is valid for
an arbitrary number of open channels, applies equally well to distinguishable and
indistinguishable particles and provides the full scattering wave function.
To obtain the full solution to the scattering problem, we essentially have to find a connection of 
the 1D scattering amplitude $f_{1D}$ and the physical $K$-Matrix $K_{oo}^{1D,phys}$. This is done by
exploiting the relation
$S=\openone+\bigl(\frac{ik}{2\pi}\bigr)^\frac{d-1}{2}\hat{f}$, \cite{lupu1998quantum}
where $S$ denotes the scattering matrix of the system and
$\hat{f}$ is an integral operator \cite{landau1977quantum} averaging the scattering amplitude over
the solid angle. It relates the scattering amplitude and the scattering matrix in arbitrary
dimensions $d$. For our
purpose we consider the special case $d=1$ and obtain, by evaluating the integral to be two times
the scattering amplitude, an expression for the 1D scattering matrix, given by
\begin{equation}
S_{1D}=\openone+2f_{1D},
\label{eq:rel_s_f1D}
\end{equation}
which, in particular differs from the standard textbook result for the 3D relation by the absence of
an additional momentum dependence compensating for the physical dimension of the 3D scattering
amplitude. Using now the Cayley transform $S=(1+iK)(1-iK)^{-1}$ 
we obtain the 1D scattering amplitude in terms of the physical $K_{oo}^{1D,phys}$
\begin{equation}
f_{1D}=\frac{i K_{oo}^{1D,phys}}{1-iK_{oo}^{1D,phys}},
\label{eq:1d_scattering_amplitude}
\end{equation}
and correspondingly the well-known expression for the transmission amplitude
\begin{equation}
T=|1+f_{1D}|^2=\frac{1}{1+(K_{oo}^{1D,phys})^2}
\label{eq:transmission_coefficient}
\end{equation}

Note that Eqs. \eqref{eq:1d_scattering_amplitude} and \eqref{eq:transmission_coefficient} hold
regardless of the particle exchange symmetry. The bosonic or fermionic character of the particles is
already embedded in the $K^{1D, \rm{phys}}_{oo}$ matrix. Furthermore, we remark that this derivation
is not restricted to quasi-1D Hamiltonians and can be equally applied to Hamiltonians of
arbitrary dimensions connecting the relevant physical $K$ matrix with the corresponding scattering
amplitudes and cross-sections.\par
In addition, let us give some general remarks on Eq.
\eqref{eq:transmission_coefficient} concerning the two significant values the transmission
coefficient can take, namely $T=0$ which especially addresses the CIR, and, $T=1$ corresponding to
total transparency characterizing in particular the absence of a back-scattering process.
As the general form of the transmission
coefficient is given by Eq. \eqref{eq:transmission_coefficient}, we immediately observe, that the
zeros of the transmission are in one-to-one correspondence with a diverging physical $K$-Matrix,
while the unit values
appear for a vanishing $K_{oo}^{1D,phys}$. The inspection of Eqs. \eqref{eq:k_oo_double} and
\eqref{eq:k_oo_single} shows, as the physical $K$-Matrix is in both cases clearly separated in
numerator and denominator, that a zero value and divergence of the physical $K$-Matrix can only be
obtained by a root of the numerator and denominator, respectively. The case of a diverging numerator
can be excluded since the elements of the local frame transformation $U_{\ell 0}$ and the particular
$\mathfrak{U}_{\ell\ell^{\prime}}$'s behave regular within the range of considered energies. In the
present investigation we exclude the threshold energy $\epsilon=1$ of the first excited channel,
since over there the elements of $\mathfrak{U}_{\ell\ell}(\epsilon)$ are in general singular and
thus lead to threshold singularities \cite{landau1977quantum,hess2014prep}, arising due to the
fact that at the channel thresholds the $S$-Matrix of the system abruptly changes its dimension
since additional transitions between the open channels become available.

\section{Results and Discussion} 
\label{sec:results}
\subsection{Universal properties of $\ell$-wave CIRs}
\label{sse:universal_ell_cirs}

\begin{figure}[htp]
 \includegraphics[width=0.5\textwidth]{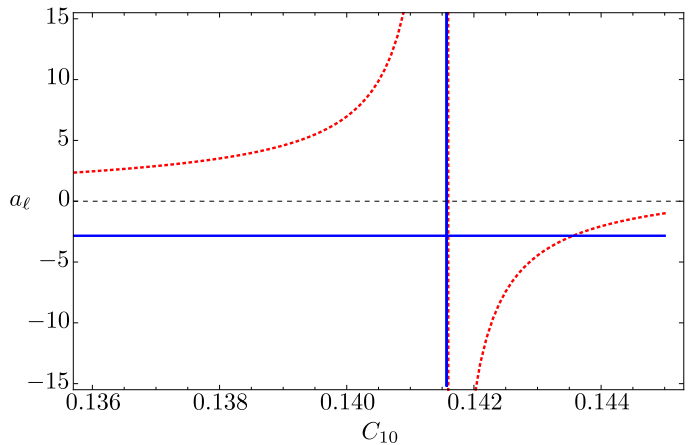}
\caption{\label{fig:freespace_sg_resonance}(Color online) The $s$-wave scattering length $a_{0}$
(red dashed line) and the $g$-wave scattering length $a_{4}$ (blue solid line) are shown versus the
$C_{10}$ parameter. It is observed, that both scattering lengths are diverging at the same values of
$C_ {10}$ meaning that the $s$- and $g$-wave bound and quasi-bound states appear simultaneously at
the threshold.}
\end{figure}

In the following we will focus on the universal properties of the $\ell$-wave CIRs. Since our main
concern is the description of neutral alkaline atomic collisions in the presence of a waveguide the following analysis
depends on the van der Waals tail of the interatomic interactions. From the perspective of
free-space collisions Gao \cite{gao2000} developed an angular
insensitive quantum defect theory focusing on pair collisions under the influence of
a van der Waals potential tail and under the additional assumption, that the
interatomic potential is sufficiently deep, i.e. supports many bound states. Investigating the universal aspects of such collisional systems it
was shown that the resonant structure of different $\ell$ partial waves possesses a periodic character
with respect to the angular momentum. More specifically, it was shown that when an $\ell$-wave (quasi-)bound state
crosses the threshold simultaneously an $\ell+4$-wave (quasi-)bound state crosses it as
well. This remarkable property is demonstrated in Fig. \ref{fig:freespace_sg_resonance} where in a transparent way we show that
$s$- (red dashed line) and $g$-wave (blue solid line) scattering lengths are diverging
simultaneously. The same holds for $p$- and $h$-wave, $d$- and $i$-wave, $f$- and $k$-wave,
respectively.\par
However, except from tuning a CIR by changing the transversal confinement frequency, still, the most common
technique is the variation of the free-space scattering length by means of a magnetic
Feshbach resonance. Therefore, in order to properly take the multi-channel nature of this phenomenon
into account, Gao developed a multi channel quantum defect theory \cite{gao2011} to describe higher
partial wave Feshbach resonances. The drawback of this theory is the fact, that the channel-closing
procedure leads to an effective short range $K$-Matrix which does depend on the angular momentum quantum
number $\ell$. Therefore, the necessary $\ell$-independence of the short range $K$-Matrix, needed to
derive the $\Delta\ell=4$ periodicity is not given and therefore we do not in general expect that
$\ell$-wave bound states around threshold appear in sets. But, as also shown by Gao in the same
work, the effective short range $K$-Matrix can again be treated as being independent of $\ell$ in
the case of broad Feshbach resonances, which are within his theory described by having a large parameter
$\zeta_{\rm res}$. This quantity is similar to the strength parameter $s_{\rm res}$ defined earlier
by Chin \emph{et al.} \cite{chin2010}. In particular it is shown, that $\zeta_{\rm res}\gg 1$ holds for
$\leftidx{^6}\rm Li$ and $\leftidx{^{133}}\rm Cs$ in various preparations.\par
Considering the same set up within a waveguide the transversal confinement results into a coupling
among all the partial waves. In this case a particular CIR possesses a specific $\ell$-wave
character when the corresponding partial wave dominates over the remainder or in other words an
$\ell$-wave CIR occurs in the vicinity of an $\ell$ wave free-space resonance. Therefore, the
interplay of the confinement with the $\ell+4$ periodicity of a deep van der Waals potential yields
four universal $\ell$-wave CIRs, namely $s,p,d,f$, overwhelming the corresponding contributions of
$\ell+4$ partial waves. This simply means that there are two $\ell$-wave CIRs for bosonic collisions
and two more for the fermionic ones.\par
As Gao's findings on the $\ell+4$ periodicity crucially depend on the assumption that all
(quasi-)bound states must be close to the threshold of the interatomic potential, we express the
energy scale associated with this potential $\epsilon_{\rm{sh}}$, introduced in \cite{gao2000}, in terms
of the confinement length scale, yielding:
\begin{equation} 
\epsilon_{\rm{sh}}=\frac{(\epsilon+1/2)\hbar}{4\mu}\;\Bigl(\frac{\beta_{6}}{a_{\perp}}\Bigr)^2,
\label{eq:shortrange_energyscale}
\end{equation}
which shows, that the length scale separation ($\beta_{6}\ll a_{\perp}$) also separates the energy
scale of the waveguide, given by $\epsilon$ and the interatomic energy scale, given by
$\epsilon_{\rm{sh}}$. Thus, we can safely assume to be in the regime nearby the threshold of the
interatomic potential while still being allowed to consider colliding energies $\epsilon\sim 1$, and thus
be well above the threshold of the transversal ground state.\par
Another universal aspect of the $\ell$-wave CIRs is the following: when a $s$ or a $p$-wave CIR
occurs, their corresponding couplings to $d$ or $f$ partial waves, respectively, can be neglected. This
is permitted since in general $d$ and $f$-wave scattering lengths are practically zero in the
vicinity of $s$- and a $p$-wave free-space resonances, respectively. Contrary to the previous case, when a $d$- or an $f$-wave
CIR occurs their corresponding couplings to the $s$- or $p$-wave can not be neglected. This occurs
due to the fact that for energies ranging in the interval $\epsilon\in [0,1)$,
 in the vicinity of a $d$-wave CIR the $s$-wave scattering length has the value
$a_{0_{bg}}\approx 2\pi/(\Gamma(1/4)^2)\beta_6\approx 0.48\;\beta_6$ and, similarly, in the vicinity of an $f$-wave CIR the
$p$-wave scattering length has the value $a_{1_{bg}}\approx -0.45\;\beta_6$, where the corresponding
analytical values are taken from \cite{gao2000}.

\subsection{$s$- and $p$-wave energy dependent CIRs}
\label{sub:sp_eCIR}
\begin{figure}[htp]
 \includegraphics[width=0.5\textwidth]{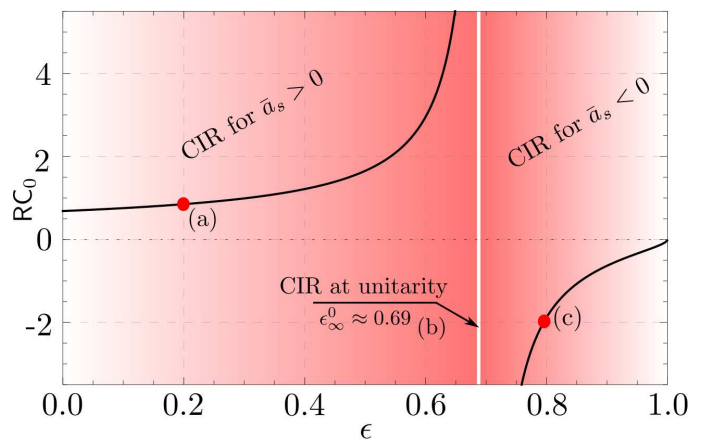}
\caption{\label{fig:bosons_rescon}(Color online) The coefficient $\mathsf{RC}_{\ell}(\epsilon)$
(black solid lines) from 
Eq.\eqref{eq:rescondition_single} versus the dimensionless energy $\epsilon$ is shown for the case
of $\ell=0$. (a) to (c) refer to the values of energy for which the transmission coefficient
$T$ is shown versus the scattering length in Fig. \ref{fig:bose_ecirs}. The shading indicates the
magnitude of the scaled scattering length ($\bar{a}_0$) that has to be met fulfilling the equality of
the resonance condition Eq. \eqref{eq:rescondition_single_real}, while the white line represents the
pole in the resonance condition after which the CIR occurs with a negative sign in scattering length.}
\end{figure}
\begin{figure}[htp]
 \includegraphics[width=0.5\textwidth]{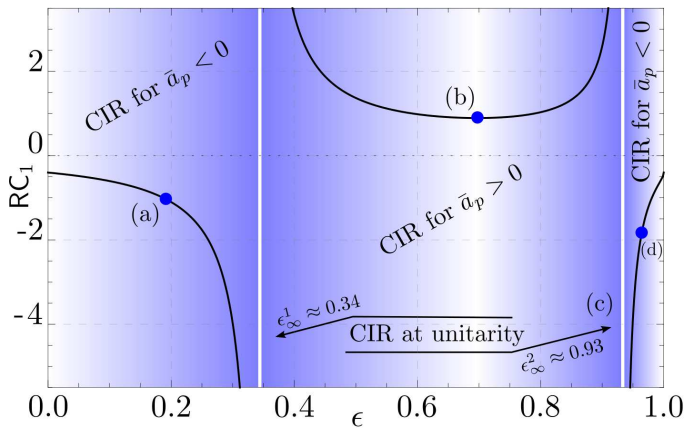}
\caption{\label{fig:fermions_rescon}(Color online) Analogous to Fig. \ref{fig:bosons_rescon}, here
$\mathsf{RC}_{1}(\epsilon)$ is shown.
 The panels (a) to (d) correspond to the values of energy for which
the transmission coefficient $T$ is shown in Fig. \ref{fig:fermi_ecirs}. Note, that this time the
resonance condition has two poles resulting in a twofold sign change in the scattering length.}
\end{figure}
For now, we turn our attention to the simplest cases where CIRs occur, i.e. $s$- and $p$-wave interactions
for bosons and spin-polarized fermions, respectively. In this case, the resonant collisions within
the waveguide result in $s$- and $p$-wave CIRs. A fundamental property of this particular type of
CIR is that they are solely characterized by one partial wave and hence are well described
within the single partial wave approach.  This is due to the fact that in the vicinity of these
CIRs all higher partial waves have practically a vanishing scattering length, namely $a_\ell\approx
0$.\par
In general, in the single partial wave picture, we obtain, by solving $\det(1-iK_{cc})=0$ for the
scattering length the \emph{resonance condition}
\begin{equation}
\bar{a}_{\ell}(\epsilon)=\mathsf{RC}_{\ell}(\epsilon)
\label{eq:rescondition_single_real}
\end{equation}
for a specific $\ell$-wave CIR, where $\mathsf{RC}_{\ell}(\epsilon)$ is given by:
\begin{equation}
  \mathsf{RC}_{\ell}(\epsilon)=\frac{-1}{2\sqrt{\epsilon+1/2}}\times
\sqrt[2\ell+1]{\frac{1}{i
\mathfrak{U}_{\ell\ell}(\epsilon)}}
  \label{eq:rescondition_single}
\end{equation}
The scaled scattering length $\bar{a}_{\ell}(\epsilon)=a_{\ell}(\epsilon)/a_{\perp}$, introduced
above, is defined as the energy dependent
$\ell$-wave free-space scattering length divided by $a_{\perp}$, whereas in turn, the energy
dependent scattering length is as usual given by: $a_{\ell}^{2\ell+1}=-\tan\delta_{\ell}/k^{2\ell+1}$.
For $\ell=0$ and
$\epsilon=0$, this relation reduces to Olshanii's result \cite{olshanii1998}, while the
corresponding case of $\epsilon=0$ and $\ell=1$ describes the $p$-wave CIR found by Granger \emph{et al.} in
\cite{granger2004}.\par

\begin{figure*}[t]
 \includegraphics[width=\textwidth]{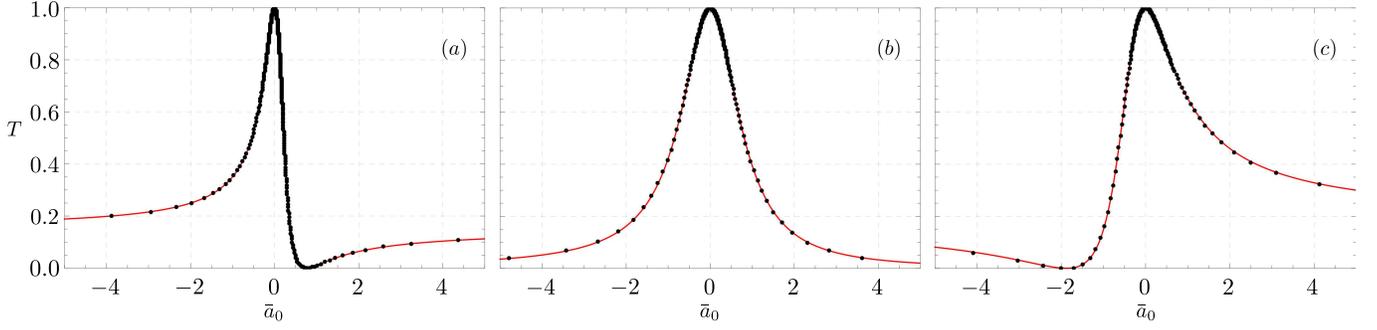}
\caption{\label{fig:bose_ecirs}The analytically calculated transmission coefficient (solid lines) is
plotted versus the scaled $s$-wave scattering length $\bar{a}_{0}$ for energies $\epsilon=0.1,0.69,0.8$ from left to right.
Corresponding to the energy values indicated by the red dots in Fig. \ref{fig:bosons_rescon}. The
black dots indicate numerical values for the transmission coefficient included for comparison.}
\end{figure*}

\begin{figure*}[t]
 \includegraphics[width=\textwidth]{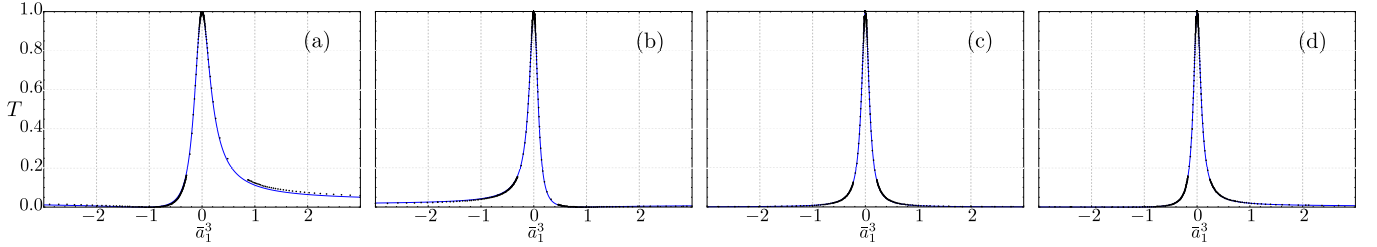}
\caption{\label{fig:fermi_ecirs}Same as Fig. \ref{fig:bose_ecirs}, but this time for
$\ell=1$ and energies $\epsilon=0.2,0.7,0.93,0.97$, from left to right, as indicated in Fig.
\ref{fig:fermions_rescon}.}
\end{figure*}

For the case of $s$-wave CIRs, we show in Fig. \ref{fig:bosons_rescon} the coefficients
$\mathsf{RC}_{0}$ versus the dimensionless energy $\epsilon$ depicted as black solid lines, while
the red shading illustrates the magnitude, which $\bar{a}_{0}(\epsilon)$ has to have at a certain
energy in order to form a CIR. Over there we readily read off the familiar value of $\bar{a}_{0}(0)=
-1/\zeta(1/2)\approx 0.68$ at $\epsilon=0$. Now, as we move
on to higher energies we see that the scaled scattering length also has to increase in order to
become resonant until, at an energy $\epsilon_{\infty}^{0}\approx 0.69$, the resonance coefficient
$\mathsf{RC}_{0}$ becomes
infinite and thus the (scaled) scattering length must be close to unitarity in order to satisfy Eq.
\eqref{eq:rescondition_single_real}.\par

Similarly to the $s$-wave CIRs, the corresponding coefficients $\mathsf{RC}_{1}$ for the
$p$-wave CIRs are shown by the black solid line in Fig. \ref{fig:fermions_rescon}, while the blue
shading this time corresponds to the magnitude of $\bar{a}_1$ in order to meet the resonance condition
Eq. \eqref{eq:rescondition_single_real}. We observe that here a similar
behavior is present. This time departing at $\epsilon=0$ with a small negative value, the resonance
coefficients monotonically decreases until they reach $-\infty$ at $\epsilon^1_{\infty}\approx 0.34$. In that sense, the main
difference between the resonance conditions for $\ell=1$ and $\ell=0$ is,
that the coefficient $\mathsf{RC}_{\ell}$ for $p$-waves has two poles, whereas the bosonic counterpart only
exhibits one pole, respectively. However, common to both coefficients is the fact, that the poles are of
odd order, or, equivalently, are related to sign changes in the resonance condition, implying that for both
partial waves the CIR can occur for both signs of the scaled scattering length. To be more specific,
for $\ell=0$ the CIR can appear for energies larger $\epsilon^0_{\infty}$ if and only if
$\bar{a}_0<0$, while for $\epsilon<\epsilon^{0}_{\infty}$, the CIR can occur only for values of
$\bar{a}_{0}$ being positive. The fermionic case of $\ell=1$ allows for positive (CIR) values of
$\bar{a}_1$ if and only if the energy lies in the interval
$\epsilon^1_{\infty}<\epsilon<\epsilon^2_{\infty}\approx 0.93$, while CIRs with a negative value of
$\bar{a}_{1}$ can occur for the remaining energies, namely $\epsilon<\epsilon^{1}_{\infty}$ and
$\epsilon>\epsilon^{2}_{\infty}$. This is in particular illustrated in Figs.
\ref{fig:bose_ecirs} (a) and (c) for bosons and Figs. \ref{fig:fermi_ecirs} (a),(b) and (d) for
fermions, respectively. There, the transmission coefficient $T$ is shown versus the corresponding scaled
scattering length $\bar{a}_{\ell}$ at different energies, which are also labeled by the
corresponding letters in Figs. \ref{fig:bosons_rescon} and \ref{fig:fermions_rescon}, respectively.
In these transmission spectra the CIRs, identified by a vanishing transmission, appear at the values given according to Eq.
\eqref{eq:rescondition_single}. We also observe, that the values of $T=1$, i.e. the total
transparency, is always located at $\bar{a}_{\ell}=0$, as it is expected by our earlier discussion
in the last paragraph of Sec. \ref{sec:observables} on the vanishing of the physical $K$-Matrix. In the case of a
single partial wave a vanishing numerator of the $K$-Matrix can only be achieved by a vanishing
scattering length, i.e. by the absence of the free-space interactions.\par
Note the black dots in Figs. \ref{fig:bose_ecirs} and \ref{fig:fermi_ecirs}, which indicate
\emph{ab initio} numerical simulations based on \cite{melezhik2012} where we solve directly the
Hamiltonian in Eq. \eqref{eq:Hamiltonian}. The numerical simulations are in excellent agreement with the corresponding
analytical calculations, namely red and blue solid lines in Figs. \ref{fig:bose_ecirs} and
\ref{fig:fermi_ecirs}, respectively.
As seen in Figs. \ref{fig:bose_ecirs} (b) and \ref{fig:fermi_ecirs} (c) the situation changes, when
considering the transmission coefficient at the particular values of energy $\epsilon^i_{\infty}$, where
$\mathsf{RC}_{\ell}(\epsilon)$ diverges. There, in particular we recognize, that the asymmetric Fano
line-shape, typical for a Fano-Feshbach resonance, is absent and instead a symmetric Lorenzian shape
of the transmission coefficient is observed.
An explanation of this effect is given by the fact, that the elements of
$\mathfrak{U}_{\ell\ell^\prime}(\epsilon)$ by construction describe the coupling of the bound states
supported by all closed channels to the open via a particular $\ell$-wave. Therefore, at the
particular values $\epsilon^{i}_{\infty}$, $i=1,2,3$, the corresponding elements vanish, yielding the decoupling of the closed channel
bound states from the continuum of the open one. In other words, this means that the pair of atoms experience an
\emph{effective} free-space collision within the waveguide, where resonant scattering occurs for
$\bar{a}_0=\infty$ as in free-space.\par
One way to describe the transition between the regime where a CIR is
present and its absence is most conveniently done by introducing the Fano $q$-parameter \cite{fano1961qparameter},
which is originally defined as the ratio between the transition probabilities to the discrete state and to the continuum.
Following this nomenclature, the symmetric line-shape is obtained when the
transition to the continuum tends to zero, namely the coupling between closed channel bound state and the
open channel continuum vanishes, and hence $q$ diverges. For a general $\ell$-wave confinement-induced
processes, we define the $\ell$-dependent $q$-parameter $q_{\ell}$ to be:
\begin{equation}
q_{\ell}:=-(\mathsf{RC}_{\ell})^{2\ell+1}
\label{eq:fano_q_parameter}
\end{equation}
Using this parametrization, the transmission coefficient reads

\begin{equation}
T=\frac{(\bar{a}^{2\ell+1}_{\ell}-q_{\ell})^2}{(\bar{a}^{2\ell+1}_{\ell}-q_{\ell})^2+(q_{\ell}\Delta_{\ell}U_{\ell
0}^2)^2}.
\label{eq:transmission_q_l}
\end{equation}

Taking now the limit $\epsilon\rightarrow \epsilon_{\infty}^{i}$, the $q$-parameter $q_{\ell}$
diverges and we end up with:
\begin{equation}
T_{\infty}= \frac{\epsilon^{0}_{\infty}}{\epsilon^{0}_{\infty}+(\bar{a}_{0})^2},
\label{eq:lorenz_bosons}
\end{equation}
as the expression for the transmission coefficient in the case of $\ell=0$, describing the Lorenzian
line shape which is solely parameterized in terms of the scaled scattering length $\bar{a}_0$.
Similarly, for the case of $\ell=1$ we obtain:
\begin{equation}
T_{\infty}=
\frac{(\epsilon^{j}_{\infty})^{-1}}{(\epsilon^{j}_{\infty})^{-1}+144(\bar{a}_{1}^{3})^2},
\label{eq:lorenz_fermions}
\end{equation}
where, in Eq. \eqref{eq:lorenz_fermions}, $\epsilon_{\infty}^j$ for $j=1,2$ refer to the energies at
the poles in the coefficients $\mathsf{RC}_{1}$.
Again, this transmission coefficient is parameterized by the scaled scattering length 
$\bar{a}_1$.\par
As we mentioned above at energies $\epsilon=\epsilon^i_{\infty}$ the resonant collisions occur at
$\bar{a}_\ell \to \infty$ yielding transmission blockade, i.e. $T=0$. In order to firmly address
this point we consider that the corresponding scattering lengths are deeply in the unitarity regime.
Then  for a single partial wave $\ell$ the general form of the transmission coefficient $T$ given in
Eq. \eqref{eq:transmission_coefficient} for scattering lengths at unitarity reduces to
$T_{\rm{unitarity}}$ according to the following relation:
\begin{equation}
T_{\rm{unitarity}}=\lim_{\Delta_{\ell}\rightarrow\infty}T=\frac{\mathfrak{U}_{\ell\ell}^2(\epsilon)}{\mathfrak{U}_{\ell\ell}^2(\epsilon)-U_{\ell
0}^4(\epsilon)},
\label{eq:transmission_unitary}
\end{equation}
\begin{figure}
\includegraphics[width=0.4\textwidth]{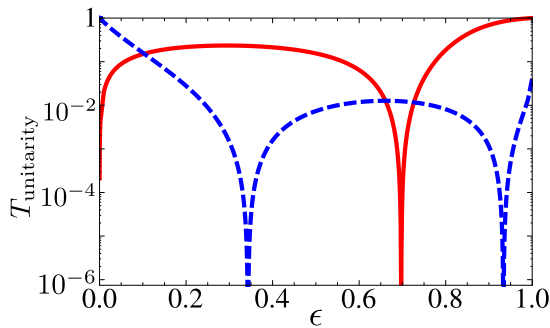}
\caption{\label{fig:unitary_transmission}(Color online) The transmission coefficient
$T_{\rm{unitary}}$ at unitarity
is shown versus the channel normalized energy $\epsilon$ for $\ell=0$ and $\ell=1$, solid and dashed
curve, respectively. Note that the minima indeed appear at $\epsilon^0_{\infty}$ for $\ell=0$ and at
$\epsilon_{\infty}^{1}$ and $\epsilon_{\infty}^{2}$ for $\ell=1$, as these denote the location of the poles in
Figs. \ref{fig:bosons_rescon} and \ref{fig:fermions_rescon}, respectively.}
\end{figure}

where we refer to the Appendix \ref{app:derivation_umatrix} for further details on
$\mathfrak{U}$. Eq. \eqref{eq:transmission_unitary} is depicted in Fig.
\ref{fig:unitary_transmission} both for $s$-wave (red line) and $p$-wave (blue dashed line) cases, where we plot
the transmission coefficient $T_{\rm{unitarity}}$ on a logarithmic scale as a function of energy $\epsilon$. As
expected we observe that indeed the transmission $T_{\rm{unitarity}}$ becomes zero at the energy
values $\epsilon_{\infty}^0$ for the $s$-wave case and
$\epsilon_{\infty}^1,\epsilon_{\infty}^{2}$ for the $p$-wave case.

\subsection{$d$- and $f$-wave energy dependent CIRs}
\label{sub:df_eCIR}

Let us now discuss the case of $d$- and $f$-wave CIRs where we will solely focus on
the universal properties of the extrema of the corresponding transmission coefficients. Unlikely
to the case of $s$- and $p$-wave CIRs,  $d$- and $f$-wave CIRs are strongly affected by the presence
of $s$- and $p$- partial waves, respectively. This occurs since in the vicinity of these CIRs the
$s$- and $p$-wave scattering lengths retain a non-vanishing value. Therefore, the single partial
wave approach is not valid anymore particularly for $T\approx1$.\par
First we consider the case $T=0$. Analogous to the case of a single partial wave, the \emph{resonance
condition} $\bar{a}_{\ell^{\prime}}=\mathsf{RC}_{\ell^{\prime},\ell}(\epsilon)$ in the presence of
two partial waves $\ell$ and $\ell^{\prime}$ is obtained by solving the
determinant from Eq.\eqref{eq:k_oo_double} for the corresponding scattering length, yielding the
following expression
\begin{equation}
  \mathsf{RC}_{\ell^{\prime},\ell}(\epsilon)=\frac{-1}{2\sqrt{\epsilon+1/2}}\times
\sqrt[2\ell^{\prime}+1]{
\frac{1}{
i\bigl(
\mathfrak{U}_{\ell^{\prime}\ell^{\prime}}-\alpha_{\ell}\mathfrak{U}^{2}_{\ell\ell^{\prime}}\bigr)}},
\label{eq:resonance_condition_double}
\end{equation}
where $\alpha_{\ell}=i\Delta_{\ell}/(1-i\Delta_{\ell}\mathfrak{U}_{\ell\ell})$ represents the coupling
strength of the $\ell^{\prime}$-wave to the $\ell$-wave and essentially describes how the free-space
process is affected by the closed channels, which can be seen by comparing $\alpha_{\ell}$ with Eq.
\eqref{eq:k_oo_single}. Equation \eqref{eq:resonance_condition_double} nicely shows, how the
corresponding coefficients from Eq. \eqref{eq:rescondition_single} are altered by the presence of a second partial wave.
Also note, that in the single partial $\ell^{\prime}$-wave approximation, we have $\Delta_{\ell}=0$, and hence
$\alpha_\ell=0$, Eq. \eqref{eq:resonance_condition_double} reduces to Eq. \eqref{eq:rescondition_single}.\par
As mentioned before, the $\ell$-wave character is addressed to a CIR if and only if the
corresponding partial wave dominates over all the others. Therefore, for the particular case of
$T=0$, $d$ and $f$-wave CIRs occur when the corresponding scattering lengths dominate. Hence, the
background scattering lengths, namely $s$ and $p$- wave can be regarded as minor corrections to the
positions of $d$- and $f$-wave CIRs.\par
Now, similar to the case of $T=0$ we investigate the solutions to the constraint $T=1$, i.e. total
transparency, since the numerator of the physical $K$-Matrix for coupled partial waves, given in Eq.
\eqref{eq:k_oo_double}, contains non-trivial relations between the different partial waves, as well
as couplings to the closed channels. Hence, analogous to the resonance condition of Eq.
\eqref{eq:rescondition_single_real}, we obtain the \emph{transparency condition} for a
$\ell^{\prime}$-wave dominated process, given by
\begin{equation}
\bar{a}_{\ell^{\prime}}(\epsilon)=\mathsf{TC}_{\ell^{\prime},\ell}(\epsilon),
\label{eq:transparancy_condition_real}
\end{equation}
where the coefficients $\mathsf{TC}_{\ell^{\prime},\ell}(\epsilon)$ are
\begin{widetext}
\begin{align}
\mathsf{TC}_{\ell^{\prime},\ell}(\epsilon)&=\frac{1}{2\sqrt{\epsilon+1/2}} \times \sqrt[2\ell^{\prime}+1]{ \frac{ \Delta_{\ell }U_{\ell 0}^2
}{U_{\ell^{\prime}0}^2-i\Delta_\ell(\mathfrak{U}_{\ell^{\prime}\ell^{\prime}}U_{\ell
0}^2+\mathfrak{U}_{\ell\ell}U_{\ell^{\prime}0}^2-2\mathfrak{U}_{\ell\ell^{\prime}}U_{\ell0}U_{\ell^{\prime}0})}},
\label{eq:unitcondition}
\end{align}
\end{widetext}
However, in the case of total transparency, we observe from Eq. \eqref{eq:unitcondition} that the
value of $\bar{a}_{\ell^{\prime}}$ for which $T$ becomes unity strongly depends on the corresponding
background scattering length as well as on the colliding energy $\epsilon$. Also note, that in the case of a
single partial wave, e.g. $\Delta_\ell=0$, the RHS of Eq. \eqref{eq:unitcondition} vanishes
identically and we are left with the conclusion from Sec.
\ref{sub:sp_eCIR}, that total transparency can occur only for $\bar{a}_{\ell}=0$. Hence, contrary to
$s$- and $p$- wave total transparency, here, in the case of coupled partial waves, the occurrence of
the total transparency is the immediate result of destructive interference between $s-d$ and $p-f$
partial waves for bosons or fermions, respectively. Therefore this feature of CIR corresponds to the
bosonic and fermionic  \emph{dual} CIR which has been discussed in Ref. \cite{kim2006} for
distinguishable particles.\par

\begin{figure}[t]
 \includegraphics[width=0.4\textwidth]{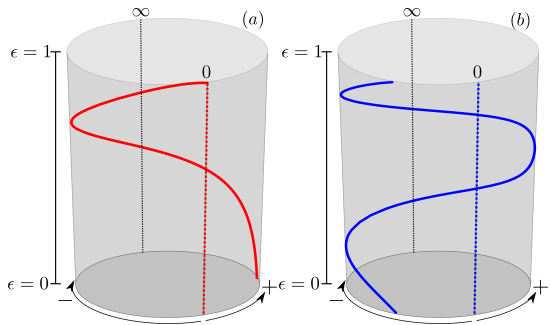}
\caption{\label{fig:cylinder_SP}(Color online) The solid lines in panels (a) and (b) depict
$\mathsf{RC}_0(\epsilon)$ and $\mathsf{RC}_1(\epsilon)$, respectively, while both dashed lines show the condition for
T=1, i.e. total transparency.}
\end{figure}

\begin{figure}[h]
 \includegraphics[width=0.4\textwidth]{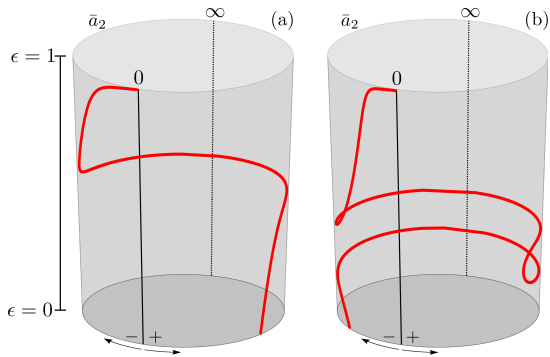}
\caption{\label{fig:cylinder_D}(Color online) For the case of $d$-wave CIR, panel (a) shows the
resonance
coefficients $\mathsf{RC_{2,0}}(\epsilon)$ versus the total colliding energy $\epsilon$. (b) shows,
also versus $\epsilon$, the transparency coefficients $\mathsf{TC}_{2,0}(\epsilon)$, which is seen to depend strongly on energy.}
\end{figure}

\begin{figure}[b]
 \includegraphics[width=0.4\textwidth]{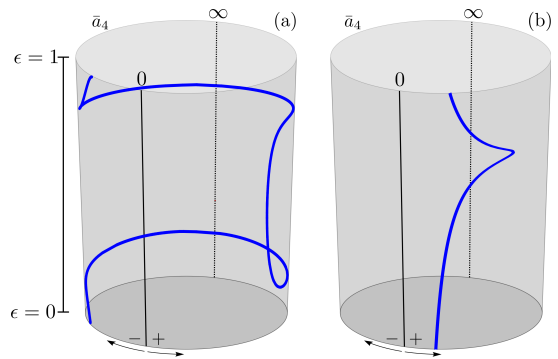}
\caption{\label{fig:cylinder_F}(Color online) 
For the case of $f$-wave CIR, analogous to Fig. \ref{fig:cylinder_D}. Panel
(a) depicts $\mathsf{RC}_{3,1}(\epsilon)$ and panel (b) $\mathsf{TC}_{3,1}(\epsilon)$. Contrary to
the bosonic counterpart shown in Fig. \ref{fig:cylinder_D} we observe from panel (b), that there is
now $f$-wave dual CIR for negative values of $\bar{a}_4$.
}
\end{figure}

Figures \ref{fig:cylinder_SP}, \ref{fig:cylinder_D} and \ref{fig:cylinder_F} present a convenient
visualization of the energy dependence of the transmission extrema for
$s$-, $p$-, $d$- and $f$-wave (dual) CIRs. This representation is achieved by stereographically projecting
on a cylinder the geometrical topos, i.e. the trajectory, of the corresponding coefficients
$\mathsf{RC}_{\ell^{\prime},\ell}$ and $\mathsf{TC}_{\ell^{\prime},\ell}$ for $T=0$ and $T=1$,
respectively. The basis of the cylinder is formed by mapping the complete range of values of the
scaled scattering length $\bar{a}_\ell$  on a circle. This particularly allows us to illustrate the values
$\bar{a}_\ell=0$ and $\bar{a}_\ell=\pm \infty$ as two anti-diametric points of the circle. In
addition perpendicularly to the plane of the circle we add the axis of the energy $\epsilon$.\par
More specifically, the panels (a) and (b) of Fig. \ref{fig:cylinder_SP} corresponds to the cases of $s$ and $p$-wave
CIRs respectively. In both panels we observe that the resonance trajectories (dashed lines)
for $T=1$ are completely straight lines on the corresponding cylindrical surfaces demonstrating in
this manner that they do not depend on energy. On the other hand the resonance trajectories for $T=0$
(solid lines) exhibit a more intricate dependence on the energy. The resonance trajectories
spiral upwards as the energy is increased illustrating in a transparent way the sensitivity of the
position of $s$ and $p$-wave CIRs which alter via the total colliding energy yielding thus CIRs even
for negative scattering lengths. This change in the sign occurs when the position of the
corresponding CIRs cross the infinity point, namely $\bar{a}_\ell=\pm \infty$. In addition in Fig.
\ref{fig:cylinder_SP} (b) we observe that the energy dependence yields a double change on the sign
of the $p$-wave scattering length as it was already shown in Figs. \ref{fig:fermions_rescon} and
\ref{fig:fermi_ecirs}.\par
Figs. \ref{fig:cylinder_D} and \ref{fig:cylinder_F} corresponds to $d$- and $f$-wave CIRs,
respectively. More specifically, panel (a) in Figs. \ref{fig:cylinder_D} and \ref{fig:cylinder_F}
refer to the cases of $d$- and $f$-wave CIRs, respectively. The resonance trajectories for $T=0$ are
denoted by red and blue solid lines for $d$- and $f$-wave CIRs, respectively. We observe that both
cases exhibit similar behavior as in the corresponding cases of $s$- and $p$-wave CIRs. This occurs
since the corresponding resonance conditions (Eqs. \eqref{eq:rescondition_single} and \eqref{eq:resonance_condition_double})
contain the same zeta function pieces attributing therefore similar behavior to $s$- and $d$-wave
CIRs or $p$- and $f$-wave CIRs.\par
Furthermore, we observe in panel (b) of Fig. \ref{fig:cylinder_D}
that the resonance trajectory for $T=1$ strongly depends
on energy yielding thus the following behavior: The position of $T=1$ spirals up initially
counterclockwise with respect to the resonance trajectory $T=0$ and evidently the corresponding
scattering length is changing sign across the point $\bar{a}_\ell=\infty$ in order to fulfill the
condition Eq. \eqref{eq:transparancy_condition_real}. Comparing this observation with the
corresponding fermionic case, i.e. the behavior of $\mathsf{TC}_{3,1}(\epsilon)$ which is shown in panel
(b) of Fig. \ref{fig:cylinder_F}, we find that the dependence on energy is not as intricate as it
is for the corresponding bosonic coefficient $\mathsf{TC}_{2,0}(\epsilon)$. This behavior is an
immediate result of the weak interference of $p$ and $f$ waves. Additionally, we
observe that there is no $f$-wave dual CIR for negative values of $\bar{a}_4$.
\par

As we mentioned above, the trajectories of Figs. \ref{fig:cylinder_SP} to \ref{fig:cylinder_F} for
$T=0$ and for $T=1$ occur from the roots of the
denominator and nominator of the physical $K$-matrix, respectively. Therefore, intersections between
the $T=0$- and $T=1$-trajectories are prohibited since this would result in an indeterminate physical
$K$-matrix. Or, in terms of physical behavior, this would yield a scenario where the transmission $T$ would be
simultaneously zero and unity at the corresponding scattering length and energy. However, we remark
that the trajectories $T=0$ and $T=1$ might approach each other at some particular values of the scaled
scattering lengths and energies and exhibit in this manner a transmission profile
where T abruptly changes from total transparency to total reflection. Hence, close to
these {\it exceptional} values the full physical $K$-matrix has to be employed and not its parts,
namely the nominator and the denominator yielding thus trajectories which do not possess crossings.

\subsection{Confinement-induced resonances and closed channel bound states}
\label{sub:cir_boundStates}

\begin{figure}[t]
 \includegraphics[width=0.4\textwidth]{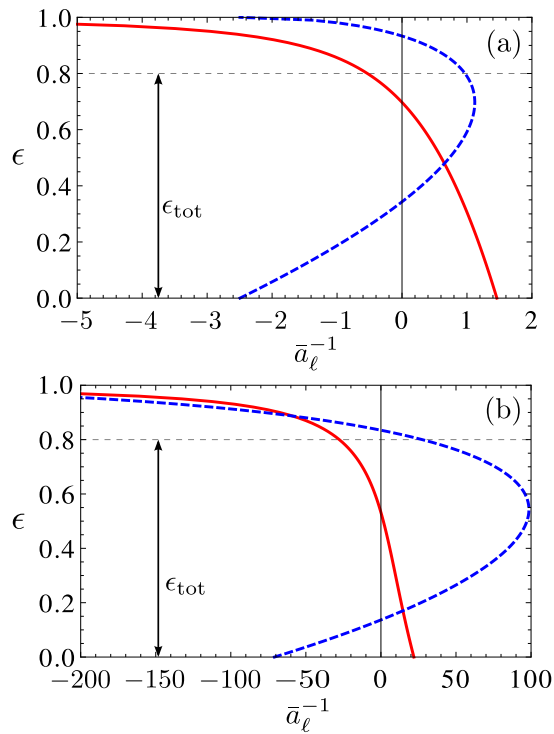}
\caption{\label{fig:sp_df_boundstates}(Color online) The eigenenergies of the closed channel
bound states of the $\ell$-wave CIRs as a function of the corresponding inverse scattering length,
namely $\bar{a}_\ell^{-1}$. (a) shows the bound states of $s$- (red solid line) and $p$-wave (blue
dashed line) CIRs and (b) depicts the bound-states of $d$- (red solid line) and $f$-wave CIRs (blue
dashed line). The horizontal dashed line illustrates the total colliding energy $\epsilon_{\rm{tot}}$.}
\end{figure}
In this subsection we address the physical interpretation of the energy dependent
$\ell$-wave CIRs. As we mentioned above this particular type of resonances fulfill a Fano-Feshbach
scenario. Therefore, a detailed analysis based on the bound eigenspectrum of the closed channels
will allow us to rigorously show that indeed an $\ell$-wave CIR can occur even when the two-body
interactions are not deep enough to sustain a (quasi-) bound state thereby going beyond
previous studies \cite{bergeman2003}.\par
In the following we will calculate the bound state eigenenergies via the roots of $\det(1-iK_{cc})$,
where the $K_{cc}$ matrix is fully energy dependent \cite{aymar1996multichannel}.
Moreover, we remark that $\det(1-iK_{cc})$ contains only the closed channel
bound states and not their couplings to the continuum of the open channel. This means that
Fig. \ref{fig:sp_df_boundstates} illustrates the bare bound states of the closed channels and not
the dressed ones. Specifically, Fig. \ref{fig:sp_df_boundstates} (a) depicts the bound states
of $s$-(red solid line) and $p$-wave CIRs (blue dashed line) as a function of their corresponding
inverse scaled scattering lengths, namely $\bar{a}_\ell^{-1}$, where the horizontal dashed line indicates
the amount of the total colliding energy $\epsilon_{\rm{tot}}$.\par
The $s$-wave case, i.e. $\ell=0$,  is in accordance with the corresponding results of
\cite{bergeman2003}. In particular we observe that for $\epsilon_{\rm{tot}}=0$ the corresponding
bound state crosses the threshold of the transversal ground state at $a_{\perp}/a_{0}=1.46\dots$, as
expected. Increasing the total colliding energy $\epsilon_{\rm{tot}}$, i.e. moving away
form the threshold of the open channel we observe that the intersection of the horizontal dashed
line with the red solid one changes its location continuously towards smaller values of
$a_{\perp}/a_{0}$ until it crosses the zero which occurs at
$\epsilon_{\rm{tot}}=\epsilon^0_{\infty}$ as it
was mentioned before (see Fig. \ref{fig:bosons_rescon}) and from that colliding energy on, the incoming wave can become
resonant with the closed channel bound state only for negative scattering lengths until the
threshold to the first excited channel is reached from below. Similarly in Fig.\ref{fig:sp_df_boundstates} (b), this general behavior is also observed
for the bound state of the $d$-wave CIRs indicated by the red solid line.\par
On the other hand, the bound states of the odd partial waves behave differently. As already observed
in the respective panel (b) of Figs. \ref{fig:cylinder_D} and \ref{fig:cylinder_F}, there are two
colliding energies for which the corresponding
trajectories for $T=0$ are crossing the value $\bar{a}_\ell=\infty$ for two different colliding energies.
This behavior is clearly demonstrated in Fig \ref{fig:sp_df_boundstates} (a) and (b) where we
observe that the corresponding blue dashed lines cross twice the value $\bar{a}_\ell^{-1}=0$,
implying, a twofold change of sign in the scaled scattering length with increasing total colliding
energy $\epsilon_{\text{tot}}$ to ensure resonant scattering within the waveguide geometry.
To conclude, Fig. \ref{fig:sp_df_boundstates} shows in a transparent way that \emph{all} $\ell$-wave CIRs render the
existence of a closed channel bound state even though the underlying two-body potential may not support a
(quasi-) bound state.

\section{Summary and Conclusions}
\label{sec:conclusions}
We have investigated the two-body scattering of bosons and spin-polarized fermions in a harmonic
waveguide, taking into account the coupling of different partial waves due to the confinement, as
well as the energy dependence of the collisional processes. Furthermore, we employ the framework of
the $K$-matrix approach presented in \cite{granger2004,giannakeas2012,giannakeas2013,zhang2013}
which we combined
with the free-space collisional theoretical framework of Gao \cite{gao1998,gao2000,gao2009}. This
permits us to obtain fully analytical results including adequate two-body interatomic interactions
which possess a van der Waals tail.\par
Throughout this work the only assumption that we considered is that the length scale associated with
the interatomic potential is smaller than the oscillator
length $a_{\perp}$, implying two regions of
different symmetry, i.e. spherical close to the center and cylindrical for large relative
distances. In addition, we present analytical formulas which provide a connection of the physical
$K$-matrix with all the relevant scattering observables, in particular to the scattering amplitudes,
and matrices. This connection provides a general form and can be applied to different confining
potentials or an arbitrary number of open channels regardless if the
collisional partners are constituted of identical or distinguishable particles.\par
In the present set up of identical particle collisions within a quasi-1D waveguide geometry we have
demonstrated the universal aspects of $\ell$-wave CIRs concluding that atomic collisions render four
types of CIRs, where two of them are attributed to bosons and the other two to spin-polarized
fermions. We show that this property arises due to the interplay of a deep van der Waals potential with the
transversal confinement. In addition, we have investigated the energy dependence of $\ell$-wave CIRs
showing that all of them possess always a closed channel bound state even if the two-body potential
is not deep enough to support a weakly or quasi-bound state. Therefore the fact that the position of
the CIRs are extremely sensitive to the total colliding energy regardless of the dimensionality of the
confining potential might be an indicator for the experimental discrepancies on the sign of the position of a
quasi-2D CIR \cite{frohlich2011,haller2010}. Furthermore, recent experiments on
$\leftidx{^{133}}\rm Cs$ have shown the existence of a $d$-wave shape resonance at $1.83$mT
\cite{chin2004}. This particular type of resonance might be utilized in order to explore the energy
dependence of higher partial wave CIRs, which from the many-body viewpoint is expected to provide a
different characteristics in the transition from Tonks-Girardeau to Super Tonks-Girardeau gas
phases, as the dual CIR, i.e. the non-interacting case is obtained at a finite value of the scaled
scattering length $\bar{a}_0$. Moreover, we observe that at some
particular colliding energies the closed channel bound states decouple from the continuum of the
open channel resulting thus into an effective free-space collision within the waveguide.
Particularly in the case of $d$ and $f$-wave collisions we observed the  dual CIR, i.e. total
transparency, for indistinguishable particles which is due to the interference of different partial
waves and exhibits a strong energy dependence.

\appendix

\section{Derivation of the $\mathfrak{U}$-Matrix}
\label{app:derivation_umatrix}
Starting with the \emph{Local Frame Transformation}:
\begin{equation}
U_{\ell n} = \frac{\sqrt{2}(-1)^{d_{0}}}{a_{\perp}}\sqrt{\frac{2\ell+1}{k
q_n}}P_{\ell}(\frac{q_n}{k}),
\end{equation}
where $d_{0}$ abbreviates $\ell/2$ or $(\ell+1)/2$ for even, respectively odd partial waves and
$P_{\ell}(\cdot)$ denotes the $\ell$-th Legendre polynomial, the focus of this subsection is the
computation of the coupling of a state $\ket{\ell}$
to a state $\ket{\ell^{\prime}}$ after undergoing a transition through the collective bound state
from the closed channels, i.e. we want to calculate
$\sum_{n=n_{o}}^{\infty}\braket{\ell|n}\braket{n|\ell^{\prime}}$, which is explicitly given by the
sum over the appropriate locale frame transformations, i.e.:
\begin{equation}
\mathfrak{U}_{\ell\ell^{\prime}} := \sum_{n=n_o}^\infty U_{\ell n}U_{\ell^{\prime} n},
\label{eq:mathfrakU_definition}
\end{equation}
where $n_{o}$ denotes the number of open channels. We note that $n_{o}$ should not be confused with
the corresponding oscillator quantum number, i.e. for a
value of $n_{o}=N$, the highest accessible oscillator mode is $\ket{N-1}$. Explicitly writing 
Eq. \eqref{eq:mathfrakU_definition} yields
\begin{align}
\sum_{n=n_o}^\infty U_{\ell n}U_{\ell^{\prime} n} &=
\frac{2\;(-1)^{\frac{\ell+\ell^{\prime}}{2}+\sigma}}{a_{\perp}^2 k}
\sqrt{(2\ell +1)(2\ell^{\prime} +1)}\times\nonumber\\
&\times \sum_{n=n_o}^\infty
\frac{P_{\ell}(\frac{q_n}{k})P_{\ell^{\prime}}(\frac{q_n}{k})}{q_n},
\label{eq:sumUab}
\end{align}
where $\sigma$ is $0,1$ in the case of even, odd partial waves $\ell$ and $\ell^{\prime}$.

A common expansion for the product of two Legendre Polynomials given by
\begin{align*}
& P_{\ell}(x)P_{\ell^{\prime}}(x)=\\
&=
\sum_{\nu=|\ell-\ell^{\prime}|}^{\ell+\ell^{\prime}}\sum_{p=0}^{\nu}2^{\nu}\begin{pmatrix}\ell&\ell^{\prime}&\nu\\0&0&0\end{pmatrix}(2\nu+1)\begin{pmatrix}\nu\\p\end{pmatrix}\begin{pmatrix}\frac{\nu+p-1}{2}\\\nu\end{pmatrix}x^{p},
\end{align*}
is used to bare the common argument $q_{n}/k$. Putting now this expansion into the sum in \eqref{eq:sumUab}, and setting
\begin{equation}
\tilde{\Gamma}(\ell,\ell^{\prime},\nu,p) =
2^{\nu}\begin{pmatrix}\ell&\ell^{\prime}&\nu\\0&0&0\end{pmatrix}(2\nu+1)\begin{pmatrix}\nu\\p\end{pmatrix}\begin{pmatrix}\frac{\nu+p-1}{2}\\\nu\end{pmatrix},
\end{equation}
where the objects depending on $\ell,\ell^{\prime}$ and $\nu$ denote the Wigner
$3J$-symbols familiar from the Clebsh-Gordon coefficients,
one ends up at
\begin{equation}
\sum_{n=n_o}^\infty
\frac{P_{\ell}(\frac{q_n}{k})P_{\ell^{\prime}}(\frac{q_n}{k})}{q_n}=\sum_{\nu=|\ell-\ell^{\prime}|}^{\ell+\ell^{\prime}}\sum_{p=0}^{\nu}\frac{\tilde{\Gamma}(\ell,\ell^{\prime},\nu,p)}{k^{\nu}}\sum_{n=n_o}^{\infty}q_n^{p-1}
\label{eq:legendreproductsum}
\end{equation}
Using now the general formula
\begin{equation}
\sum_{n=n_{o}}^\infty q_{n}^{j} = \Bigl(\frac{2
i}{a_{\perp}}\Bigr)^j\zeta(-\frac{j}{2},n_{o}-\epsilon),
\label{eq:sum_q_n^}
 \end{equation}
where the RHS of this equation is regarded as the regularized value of the diverging series on the left.
Inserting now Eqs. \eqref{eq:sum_q_n^} and \eqref{eq:legendreproductsum} in Eq. \eqref{eq:sumUab} , while
also replacing $a_{\perp}k\mapsto2\sqrt{\epsilon+1/2}$, one ends up with
\begin{align}
  \mathfrak{U}_{\ell\ell^{\prime}} =& \sum_{n=n_o}^{\infty}U_{\ell
  n}U_{\ell^{\prime} n}
=\nonumber \\
 =&
 (-1)^{\frac{\ell+\ell^{\prime}}{2}+\sigma}\sqrt{(2\ell+1)(2\ell^{\prime}+1)}\times \nonumber\\
 &
 \sum_{\nu=|\ell-\ell^{\prime}|}^{\ell+\ell^{\prime}}\sum_{p=0}^{\nu}\frac{\Gamma(\ell,\ell^{\prime},\nu,p)}{(\epsilon+\frac{1}{2})^{\frac{p+1}{2}}}\zeta(-\frac{p-1}{2},n_{o}-\epsilon),
\end{align}
where $\tilde\Gamma$ is redefined such that it includes the appearing powers of the imaginary unit and an
additional factor of $2^{-1}$ from the replacement made above, thus yielding
\begin{equation}
  \Gamma(\ell,\ell^{\prime},\nu,p)=i^{p-1}\;2^{-1}\;\tilde\Gamma(\ell,\ell^{\prime},\nu,p)
\end{equation}

\bibliography{litecir}

\begin{thebibliography}{53}%
\makeatletter
\providecommand \@ifxundefined [1]{%
 \@ifx{#1\undefined}
}%
\providecommand \@ifnum [1]{%
 \ifnum #1\expandafter \@firstoftwo
 \else \expandafter \@secondoftwo
 \fi
}%
\providecommand \@ifx [1]{%
 \ifx #1\expandafter \@firstoftwo
 \else \expandafter \@secondoftwo
 \fi
}%
\providecommand \natexlab [1]{#1}%
\providecommand \enquote  [1]{``#1''}%
\providecommand \bibnamefont  [1]{#1}%
\providecommand \bibfnamefont [1]{#1}%
\providecommand \citenamefont [1]{#1}%
\providecommand \href@noop [0]{\@secondoftwo}%
\providecommand \href [0]{\begingroup \@sanitize@url \@href}%
\providecommand \@href[1]{\@@startlink{#1}\@@href}%
\providecommand \@@href[1]{\endgroup#1\@@endlink}%
\providecommand \@sanitize@url [0]{\catcode `\\12\catcode `\$12\catcode
  `\&12\catcode `\#12\catcode `\^12\catcode `\_12\catcode `\%12\relax}%
\providecommand \@@startlink[1]{}%
\providecommand \@@endlink[0]{}%
\providecommand \url  [0]{\begingroup\@sanitize@url \@url }%
\providecommand \@url [1]{\endgroup\@href {#1}{\urlprefix }}%
\providecommand \urlprefix  [0]{URL }%
\providecommand \Eprint [0]{\href }%
\providecommand \doibase [0]{http://dx.doi.org/}%
\providecommand \selectlanguage [0]{\@gobble}%
\providecommand \bibinfo  [0]{\@secondoftwo}%
\providecommand \bibfield  [0]{\@secondoftwo}%
\providecommand \translation [1]{[#1]}%
\providecommand \BibitemOpen [0]{}%
\providecommand \bibitemStop [0]{}%
\providecommand \bibitemNoStop [0]{.\EOS\space}%
\providecommand \EOS [0]{\spacefactor3000\relax}%
\providecommand \BibitemShut  [1]{\csname bibitem#1\endcsname}%
\let\auto@bib@innerbib\@empty
\bibitem [{\citenamefont {Kinoshita}\ \emph {et~al.}(2004)\citenamefont
  {Kinoshita}, \citenamefont {Wenger},\ and\ \citenamefont
  {Weiss}}]{kinoshita2004}%
  \BibitemOpen
  \bibfield  {author} {\bibinfo {author} {\bibfnamefont {T.}~\bibnamefont
  {Kinoshita}}, \bibinfo {author} {\bibfnamefont {T.}~\bibnamefont {Wenger}}, \
  and\ \bibinfo {author} {\bibfnamefont {D.~S.}\ \bibnamefont {Weiss}},\
  }\href@noop {} {\bibfield  {journal} {\bibinfo  {journal} {Science}\ }\textbf
  {\bibinfo {volume} {305}},\ \bibinfo {pages} {1125} (\bibinfo {year}
  {2004})}\BibitemShut {NoStop}%
\bibitem [{\citenamefont {Paredes}\ \emph {et~al.}(2004)\citenamefont
  {Paredes}, \citenamefont {Widera}, \citenamefont {Murg}, \citenamefont
  {Mandel}, \citenamefont {F\"{o}lling}, \citenamefont {Cirac}, \citenamefont
  {Shlyapnikov}, \citenamefont {H\"{a}nsch},\ and\ \citenamefont
  {Bloch}}]{paredes2004}%
  \BibitemOpen
  \bibfield  {author} {\bibinfo {author} {\bibfnamefont {B.}~\bibnamefont
  {Paredes}}, \bibinfo {author} {\bibfnamefont {A.}~\bibnamefont {Widera}},
  \bibinfo {author} {\bibfnamefont {V.}~\bibnamefont {Murg}}, \bibinfo {author}
  {\bibfnamefont {O.}~\bibnamefont {Mandel}}, \bibinfo {author} {\bibfnamefont
  {S.}~\bibnamefont {F\"{o}lling}}, \bibinfo {author} {\bibfnamefont
  {I.}~\bibnamefont {Cirac}}, \bibinfo {author} {\bibfnamefont {G.~V.}\
  \bibnamefont {Shlyapnikov}}, \bibinfo {author} {\bibfnamefont {T.~W.}\
  \bibnamefont {H\"{a}nsch}}, \ and\ \bibinfo {author} {\bibfnamefont
  {I.}~\bibnamefont {Bloch}},\ }\href@noop {} {\bibfield  {journal} {\bibinfo
  {journal} {Nature}\ }\textbf {\bibinfo {volume} {429}},\ \bibinfo {pages}
  {277} (\bibinfo {year} {2004})}\BibitemShut {NoStop}%
\bibitem [{\citenamefont {Haller}\ \emph {et~al.}(2009)\citenamefont {Haller},
  \citenamefont {Gustavsson}, \citenamefont {Mark}, \citenamefont {Danzl},
  \citenamefont {Hart}, \citenamefont {Pupillo},\ and\ \citenamefont
  {N\"{a}gerl}}]{haller2009}%
  \BibitemOpen
  \bibfield  {author} {\bibinfo {author} {\bibfnamefont {E.}~\bibnamefont
  {Haller}}, \bibinfo {author} {\bibfnamefont {M.}~\bibnamefont {Gustavsson}},
  \bibinfo {author} {\bibfnamefont {M.~J.}\ \bibnamefont {Mark}}, \bibinfo
  {author} {\bibfnamefont {J.~G.}\ \bibnamefont {Danzl}}, \bibinfo {author}
  {\bibfnamefont {R.}~\bibnamefont {Hart}}, \bibinfo {author} {\bibfnamefont
  {G.}~\bibnamefont {Pupillo}}, \ and\ \bibinfo {author} {\bibfnamefont
  {H.~C.}\ \bibnamefont {N\"{a}gerl}},\ }\href@noop {} {\bibfield  {journal}
  {\bibinfo  {journal} {Science}\ }\textbf {\bibinfo {volume} {325}},\ \bibinfo
  {pages} {1224} (\bibinfo {year} {2009})}\BibitemShut {NoStop}%
\bibitem [{\citenamefont {Chin}\ \emph {et~al.}(2010)\citenamefont {Chin},
  \citenamefont {Grimm}, \citenamefont {Julienne},\ and\ \citenamefont
  {Tiesinga}}]{chin2010}%
  \BibitemOpen
  \bibfield  {author} {\bibinfo {author} {\bibfnamefont {C.}~\bibnamefont
  {Chin}}, \bibinfo {author} {\bibfnamefont {R.}~\bibnamefont {Grimm}},
  \bibinfo {author} {\bibfnamefont {P.~S.}\ \bibnamefont {Julienne}}, \ and\
  \bibinfo {author} {\bibfnamefont {E.}~\bibnamefont {Tiesinga}},\ }\href@noop
  {} {\bibfield  {journal} {\bibinfo  {journal} {Rev. Mod. Phys.}\ }\textbf
  {\bibinfo {volume} {82}},\ \bibinfo {pages} {1225} (\bibinfo {year}
  {2010})}\BibitemShut {NoStop}%
\bibitem [{\citenamefont {Inouye}\ \emph {et~al.}(1998)\citenamefont {Inouye},
  \citenamefont {Andrews}, \citenamefont {Stenger}, \citenamefont {Miesner},
  \citenamefont {Stamper-Kurn},\ and\ \citenamefont {Ketterle}}]{inouye1998}%
  \BibitemOpen
  \bibfield  {author} {\bibinfo {author} {\bibfnamefont {S.}~\bibnamefont
  {Inouye}}, \bibinfo {author} {\bibfnamefont {M.~R.}\ \bibnamefont {Andrews}},
  \bibinfo {author} {\bibfnamefont {J.}~\bibnamefont {Stenger}}, \bibinfo
  {author} {\bibfnamefont {H.~J.}\ \bibnamefont {Miesner}}, \bibinfo {author}
  {\bibfnamefont {D.~M.}\ \bibnamefont {Stamper-Kurn}}, \ and\ \bibinfo
  {author} {\bibfnamefont {W.}~\bibnamefont {Ketterle}},\ }\href@noop {}
  {\bibfield  {journal} {\bibinfo  {journal} {Nature}\ }\textbf {\bibinfo
  {volume} {392}},\ \bibinfo {pages} {151} (\bibinfo {year}
  {1998})}\BibitemShut {NoStop}%
\bibitem [{\citenamefont {K\"{o}hler}\ \emph {et~al.}(2006)\citenamefont
  {K\"{o}hler}, \citenamefont {G\'{o}ral},\ and\ \citenamefont
  {Julienne}}]{kohler2006}%
  \BibitemOpen
  \bibfield  {author} {\bibinfo {author} {\bibfnamefont {T.}~\bibnamefont
  {K\"{o}hler}}, \bibinfo {author} {\bibfnamefont {K.}~\bibnamefont
  {G\'{o}ral}}, \ and\ \bibinfo {author} {\bibfnamefont {P.~S.}\ \bibnamefont
  {Julienne}},\ }\href@noop {} {\bibfield  {journal} {\bibinfo  {journal} {Rev.
  Mod. Phys.}\ }\textbf {\bibinfo {volume} {78}},\ \bibinfo {pages} {1311}
  (\bibinfo {year} {2006})}\BibitemShut {NoStop}%
\bibitem [{\citenamefont {Olshanii}(1998)}]{olshanii1998}%
  \BibitemOpen
  \bibfield  {author} {\bibinfo {author} {\bibfnamefont {M.}~\bibnamefont
  {Olshanii}},\ }\href@noop {} {\bibfield  {journal} {\bibinfo  {journal}
  {Phys. Rev. Lett.}\ }\textbf {\bibinfo {volume} {81}},\ \bibinfo {pages}
  {938} (\bibinfo {year} {1998})}\BibitemShut {NoStop}%
\bibitem [{\citenamefont {Bergeman}\ \emph {et~al.}(2003)\citenamefont
  {Bergeman}, \citenamefont {Moore},\ and\ \citenamefont
  {Olshanii}}]{bergeman2003}%
  \BibitemOpen
  \bibfield  {author} {\bibinfo {author} {\bibfnamefont {T.}~\bibnamefont
  {Bergeman}}, \bibinfo {author} {\bibfnamefont {M.~G.}\ \bibnamefont {Moore}},
  \ and\ \bibinfo {author} {\bibfnamefont {M.}~\bibnamefont {Olshanii}},\
  }\href@noop {} {\bibfield  {journal} {\bibinfo  {journal} {Phys. Rev. Lett.}\
  }\textbf {\bibinfo {volume} {91}},\ \bibinfo {pages} {163201} (\bibinfo
  {year} {2003})}\BibitemShut {NoStop}%
\bibitem [{\citenamefont {Yurovsky}\ \emph {et~al.}(2008)\citenamefont
  {Yurovsky}, \citenamefont {Olshanii},\ and\ \citenamefont
  {Weiss}}]{yurovsky2008}%
  \BibitemOpen
  \bibfield  {author} {\bibinfo {author} {\bibfnamefont {V.~A.}\ \bibnamefont
  {Yurovsky}}, \bibinfo {author} {\bibfnamefont {M.}~\bibnamefont {Olshanii}},
  \ and\ \bibinfo {author} {\bibfnamefont {D.~S.}\ \bibnamefont {Weiss}},\
  }\href@noop {} {\bibfield  {journal} {\bibinfo  {journal} {Advances In
  Atomic, Molecular, and Optical Physics}\ }\textbf {\bibinfo {volume} {55}},\
  \bibinfo {pages} {61} (\bibinfo {year} {2008})}\BibitemShut {NoStop}%
\bibitem [{\citenamefont {Dunjko}\ \emph {et~al.}(2011)\citenamefont {Dunjko},
  \citenamefont {Moore}, \citenamefont {Bergeman},\ and\ \citenamefont
  {Olshanii}}]{dunjko2011}%
  \BibitemOpen
  \bibfield  {author} {\bibinfo {author} {\bibfnamefont {V.}~\bibnamefont
  {Dunjko}}, \bibinfo {author} {\bibfnamefont {M.~G.}\ \bibnamefont {Moore}},
  \bibinfo {author} {\bibfnamefont {T.}~\bibnamefont {Bergeman}}, \ and\
  \bibinfo {author} {\bibfnamefont {M.}~\bibnamefont {Olshanii}},\ }\href@noop
  {} {\bibfield  {journal} {\bibinfo  {journal} {Advances In Atomic, Molecular,
  and Optical Physics}\ }\textbf {\bibinfo {volume} {60}},\ \bibinfo {pages}
  {461} (\bibinfo {year} {2011})}\BibitemShut {NoStop}%
\bibitem [{\citenamefont {Haller}\ \emph {et~al.}(2010)\citenamefont {Haller},
  \citenamefont {Mark}, \citenamefont {Hart}, \citenamefont {Danzl},
  \citenamefont {Reichs\"{o}llner}, \citenamefont {Melezhik}, \citenamefont
  {Schmelcher},\ and\ \citenamefont {N\"{a}gerl}}]{haller2010}%
  \BibitemOpen
  \bibfield  {author} {\bibinfo {author} {\bibfnamefont {E.}~\bibnamefont
  {Haller}}, \bibinfo {author} {\bibfnamefont {M.~J.}\ \bibnamefont {Mark}},
  \bibinfo {author} {\bibfnamefont {R.}~\bibnamefont {Hart}}, \bibinfo {author}
  {\bibfnamefont {J.~G.}\ \bibnamefont {Danzl}}, \bibinfo {author}
  {\bibfnamefont {L.}~\bibnamefont {Reichs\"{o}llner}}, \bibinfo {author}
  {\bibfnamefont {V.}~\bibnamefont {Melezhik}}, \bibinfo {author}
  {\bibfnamefont {P.}~\bibnamefont {Schmelcher}}, \ and\ \bibinfo {author}
  {\bibfnamefont {H.~C.}\ \bibnamefont {N\"{a}gerl}},\ }\href@noop {}
  {\bibfield  {journal} {\bibinfo  {journal} {Phys. Rev. Lett.}\ }\textbf
  {\bibinfo {volume} {104}},\ \bibinfo {pages} {153203} (\bibinfo {year}
  {2010})}\BibitemShut {NoStop}%
\bibitem [{\citenamefont {Sala}\ \emph {et~al.}(2013)\citenamefont {Sala},
  \citenamefont {Z\"{u}rn}, \citenamefont {Lompe}, \citenamefont {Wenz},
  \citenamefont {Murmann}, \citenamefont {Serwane}, \citenamefont {Jochim},\
  and\ \citenamefont {Saenz}}]{sala2013}%
  \BibitemOpen
  \bibfield  {author} {\bibinfo {author} {\bibfnamefont {S.}~\bibnamefont
  {Sala}}, \bibinfo {author} {\bibfnamefont {G.}~\bibnamefont {Z\"{u}rn}},
  \bibinfo {author} {\bibfnamefont {T.}~\bibnamefont {Lompe}}, \bibinfo
  {author} {\bibfnamefont {A.~N.}\ \bibnamefont {Wenz}}, \bibinfo {author}
  {\bibfnamefont {S.}~\bibnamefont {Murmann}}, \bibinfo {author} {\bibfnamefont
  {F.}~\bibnamefont {Serwane}}, \bibinfo {author} {\bibfnamefont
  {S.}~\bibnamefont {Jochim}}, \ and\ \bibinfo {author} {\bibfnamefont
  {A.}~\bibnamefont {Saenz}},\ }\href@noop {} {\bibfield  {journal} {\bibinfo
  {journal} {Phys. Rev. Lett.}\ }\textbf {\bibinfo {volume} {110}},\ \bibinfo
  {pages} {203202} (\bibinfo {year} {2013})}\BibitemShut {NoStop}%
\bibitem [{\citenamefont {Fr\"{o}hlich}\ \emph {et~al.}(2011)\citenamefont
  {Fr\"{o}hlich}, \citenamefont {Feld}, \citenamefont {Vogt}, \citenamefont
  {Koschorreck}, \citenamefont {Zwerger},\ and\ \citenamefont
  {K\"{o}hl}}]{frohlich2011}%
  \BibitemOpen
  \bibfield  {author} {\bibinfo {author} {\bibfnamefont {B.}~\bibnamefont
  {Fr\"{o}hlich}}, \bibinfo {author} {\bibfnamefont {M.}~\bibnamefont {Feld}},
  \bibinfo {author} {\bibfnamefont {E.}~\bibnamefont {Vogt}}, \bibinfo {author}
  {\bibfnamefont {M.}~\bibnamefont {Koschorreck}}, \bibinfo {author}
  {\bibfnamefont {W.}~\bibnamefont {Zwerger}}, \ and\ \bibinfo {author}
  {\bibfnamefont {M.}~\bibnamefont {K\"{o}hl}},\ }\href@noop {} {\bibfield
  {journal} {\bibinfo  {journal} {Phys. Rev. Lett.}\ }\textbf {\bibinfo
  {volume} {106}},\ \bibinfo {pages} {105301} (\bibinfo {year}
  {2011})}\BibitemShut {NoStop}%
\bibitem [{\citenamefont {G\"{u}nter}\ \emph {et~al.}(2005)\citenamefont
  {G\"{u}nter}, \citenamefont {St\"{o}ferle}, \citenamefont {Moritz},
  \citenamefont {K\"{o}hl},\ and\ \citenamefont {Esslinger}}]{gunter2005}%
  \BibitemOpen
  \bibfield  {author} {\bibinfo {author} {\bibfnamefont {K.}~\bibnamefont
  {G\"{u}nter}}, \bibinfo {author} {\bibfnamefont {T.}~\bibnamefont
  {St\"{o}ferle}}, \bibinfo {author} {\bibfnamefont {H.}~\bibnamefont
  {Moritz}}, \bibinfo {author} {\bibfnamefont {M.}~\bibnamefont {K\"{o}hl}}, \
  and\ \bibinfo {author} {\bibfnamefont {T.}~\bibnamefont {Esslinger}},\
  }\href@noop {} {\bibfield  {journal} {\bibinfo  {journal} {Phys. Rev. Lett.}\
  }\textbf {\bibinfo {volume} {95}},\ \bibinfo {pages} {230401} (\bibinfo
  {year} {2005})}\BibitemShut {NoStop}%
\bibitem [{\citenamefont {Moritz}\ \emph {et~al.}(2005)\citenamefont {Moritz},
  \citenamefont {St\"{o}ferle}, \citenamefont {G\"{u}nter}, \citenamefont
  {K\"{o}hl},\ and\ \citenamefont {Esslinger}}]{moritz2005}%
  \BibitemOpen
  \bibfield  {author} {\bibinfo {author} {\bibfnamefont {H.}~\bibnamefont
  {Moritz}}, \bibinfo {author} {\bibfnamefont {T.}~\bibnamefont
  {St\"{o}ferle}}, \bibinfo {author} {\bibfnamefont {K.}~\bibnamefont
  {G\"{u}nter}}, \bibinfo {author} {\bibfnamefont {M.}~\bibnamefont
  {K\"{o}hl}}, \ and\ \bibinfo {author} {\bibfnamefont {T.}~\bibnamefont
  {Esslinger}},\ }\href@noop {} {\bibfield  {journal} {\bibinfo  {journal}
  {Phys. Rev. Lett.}\ }\textbf {\bibinfo {volume} {94}},\ \bibinfo {pages}
  {210401} (\bibinfo {year} {2005})}\BibitemShut {NoStop}%
\bibitem [{\citenamefont {Lamporesi}\ \emph {et~al.}(2010)\citenamefont
  {Lamporesi}, \citenamefont {Catani}, \citenamefont {Barontini}, \citenamefont
  {Nishida}, \citenamefont {Inguscio},\ and\ \citenamefont
  {Minardi}}]{lamporesi2010}%
  \BibitemOpen
  \bibfield  {author} {\bibinfo {author} {\bibfnamefont {G.}~\bibnamefont
  {Lamporesi}}, \bibinfo {author} {\bibfnamefont {J.}~\bibnamefont {Catani}},
  \bibinfo {author} {\bibfnamefont {G.}~\bibnamefont {Barontini}}, \bibinfo
  {author} {\bibfnamefont {Y.}~\bibnamefont {Nishida}}, \bibinfo {author}
  {\bibfnamefont {M.}~\bibnamefont {Inguscio}}, \ and\ \bibinfo {author}
  {\bibfnamefont {F.}~\bibnamefont {Minardi}},\ }\href@noop {} {\bibfield
  {journal} {\bibinfo  {journal} {Phys. Rev. Lett.}\ }\textbf {\bibinfo
  {volume} {104}},\ \bibinfo {pages} {153202} (\bibinfo {year}
  {2010})}\BibitemShut {NoStop}%
\bibitem [{\citenamefont {Kim}\ \emph {et~al.}(2006)\citenamefont {Kim},
  \citenamefont {Melezhik},\ and\ \citenamefont {Schmelcher}}]{kim2006}%
  \BibitemOpen
  \bibfield  {author} {\bibinfo {author} {\bibfnamefont {J.~I.}\ \bibnamefont
  {Kim}}, \bibinfo {author} {\bibfnamefont {V.~S.}\ \bibnamefont {Melezhik}}, \
  and\ \bibinfo {author} {\bibfnamefont {P.}~\bibnamefont {Schmelcher}},\
  }\href@noop {} {\bibfield  {journal} {\bibinfo  {journal} {Phys. Rev. Lett.}\
  }\textbf {\bibinfo {volume} {97}},\ \bibinfo {pages} {193203} (\bibinfo
  {year} {2006})}\BibitemShut {NoStop}%
\bibitem [{\citenamefont {Granger}\ and\ \citenamefont
  {Blume}(2004)}]{granger2004}%
  \BibitemOpen
  \bibfield  {author} {\bibinfo {author} {\bibfnamefont {B.~E.}\ \bibnamefont
  {Granger}}\ and\ \bibinfo {author} {\bibfnamefont {D.}~\bibnamefont
  {Blume}},\ }\href@noop {} {\bibfield  {journal} {\bibinfo  {journal} {Phys.
  Rev. Lett.}\ }\textbf {\bibinfo {volume} {92}},\ \bibinfo {pages} {133202}
  (\bibinfo {year} {2004})}\BibitemShut {NoStop}%
\bibitem [{\citenamefont {Giannakeas}\ \emph {et~al.}(2012)\citenamefont
  {Giannakeas}, \citenamefont {Diakonos},\ and\ \citenamefont
  {Schmelcher}}]{giannakeas2012}%
  \BibitemOpen
  \bibfield  {author} {\bibinfo {author} {\bibfnamefont {P.}~\bibnamefont
  {Giannakeas}}, \bibinfo {author} {\bibfnamefont {F.~K.}\ \bibnamefont
  {Diakonos}}, \ and\ \bibinfo {author} {\bibfnamefont {P.}~\bibnamefont
  {Schmelcher}},\ }\href@noop {} {\bibfield  {journal} {\bibinfo  {journal}
  {Phys. Rev. A}\ }\textbf {\bibinfo {volume} {86}},\ \bibinfo {pages} {042703}
  (\bibinfo {year} {2012})}\BibitemShut {NoStop}%
\bibitem [{\citenamefont {Moore}\ \emph {et~al.}(2004)\citenamefont {Moore},
  \citenamefont {Bergeman},\ and\ \citenamefont {Olshanii}}]{moore2004}%
  \BibitemOpen
  \bibfield  {author} {\bibinfo {author} {\bibfnamefont {M.~G.}\ \bibnamefont
  {Moore}}, \bibinfo {author} {\bibfnamefont {T.}~\bibnamefont {Bergeman}}, \
  and\ \bibinfo {author} {\bibfnamefont {M.}~\bibnamefont {Olshanii}},\ }in\
  \href@noop {} {\emph {\bibinfo {booktitle} {Journal de Physique IV
  (Proceedings)}}},\ Vol.\ \bibinfo {volume} {116}\ (\bibinfo {organization}
  {EDP sciences},\ \bibinfo {year} {2004})\ pp.\ \bibinfo {pages}
  {69--86}\BibitemShut {NoStop}%
\bibitem [{\citenamefont {Saeidian}\ \emph {et~al.}(2008)\citenamefont
  {Saeidian}, \citenamefont {Melezhik},\ and\ \citenamefont
  {Schmelcher}}]{saeidian2008}%
  \BibitemOpen
  \bibfield  {author} {\bibinfo {author} {\bibfnamefont {S.}~\bibnamefont
  {Saeidian}}, \bibinfo {author} {\bibfnamefont {V.~S.}\ \bibnamefont
  {Melezhik}}, \ and\ \bibinfo {author} {\bibfnamefont {P.}~\bibnamefont
  {Schmelcher}},\ }\href@noop {} {\bibfield  {journal} {\bibinfo  {journal}
  {Phys. Rev. A}\ }\textbf {\bibinfo {volume} {77}},\ \bibinfo {pages} {042721}
  (\bibinfo {year} {2008})}\BibitemShut {NoStop}%
\bibitem [{\citenamefont {Saeidian}\ \emph {et~al.}(2012)\citenamefont
  {Saeidian}, \citenamefont {Melezhik},\ and\ \citenamefont
  {Schmelcher}}]{saeidian2012}%
  \BibitemOpen
  \bibfield  {author} {\bibinfo {author} {\bibfnamefont {S.}~\bibnamefont
  {Saeidian}}, \bibinfo {author} {\bibfnamefont {V.~S.}\ \bibnamefont
  {Melezhik}}, \ and\ \bibinfo {author} {\bibfnamefont {P.}~\bibnamefont
  {Schmelcher}},\ }\href@noop {} {\bibfield  {journal} {\bibinfo  {journal}
  {Phys. Rev. A}\ }\textbf {\bibinfo {volume} {86}},\ \bibinfo {pages} {062713}
  (\bibinfo {year} {2012})}\BibitemShut {NoStop}%
\bibitem [{\citenamefont {Melezhik}\ and\ \citenamefont
  {Schmelcher}(2011)}]{melezhik2011}%
  \BibitemOpen
  \bibfield  {author} {\bibinfo {author} {\bibfnamefont {V.~S.}\ \bibnamefont
  {Melezhik}}\ and\ \bibinfo {author} {\bibfnamefont {P.}~\bibnamefont
  {Schmelcher}},\ }\href@noop {} {\bibfield  {journal} {\bibinfo  {journal}
  {Phys. Rev. A}\ }\textbf {\bibinfo {volume} {84}},\ \bibinfo {pages} {042712}
  (\bibinfo {year} {2011})}\BibitemShut {NoStop}%
\bibitem [{\citenamefont {Peng}\ \emph {et~al.}(2011)\citenamefont {Peng},
  \citenamefont {Hu}, \citenamefont {Liu},\ and\ \citenamefont
  {Drummond}}]{peng2011}%
  \BibitemOpen
  \bibfield  {author} {\bibinfo {author} {\bibfnamefont {S.~G.}\ \bibnamefont
  {Peng}}, \bibinfo {author} {\bibfnamefont {H.}~\bibnamefont {Hu}}, \bibinfo
  {author} {\bibfnamefont {X.~J.}\ \bibnamefont {Liu}}, \ and\ \bibinfo
  {author} {\bibfnamefont {P.~D.}\ \bibnamefont {Drummond}},\ }\href@noop {}
  {\bibfield  {journal} {\bibinfo  {journal} {Phys. Rev. A}\ }\textbf {\bibinfo
  {volume} {84}},\ \bibinfo {pages} {043619} (\bibinfo {year}
  {2011})}\BibitemShut {NoStop}%
\bibitem [{\citenamefont {Sala}\ \emph {et~al.}(2012)\citenamefont {Sala},
  \citenamefont {Schneider},\ and\ \citenamefont {Saenz}}]{sala2012}%
  \BibitemOpen
  \bibfield  {author} {\bibinfo {author} {\bibfnamefont {S.}~\bibnamefont
  {Sala}}, \bibinfo {author} {\bibfnamefont {P.~I.}\ \bibnamefont {Schneider}},
  \ and\ \bibinfo {author} {\bibfnamefont {A.}~\bibnamefont {Saenz}},\
  }\href@noop {} {\bibfield  {journal} {\bibinfo  {journal} {Phys. Rev. Lett.}\
  }\textbf {\bibinfo {volume} {109}},\ \bibinfo {pages} {073201} (\bibinfo
  {year} {2012})}\BibitemShut {NoStop}%
\bibitem [{\citenamefont {Peano}\ \emph {et~al.}(2005)\citenamefont {Peano},
  \citenamefont {Thorwart}, \citenamefont {Mora},\ and\ \citenamefont
  {Egger}}]{peano2005}%
  \BibitemOpen
  \bibfield  {author} {\bibinfo {author} {\bibfnamefont {V.}~\bibnamefont
  {Peano}}, \bibinfo {author} {\bibfnamefont {M.}~\bibnamefont {Thorwart}},
  \bibinfo {author} {\bibfnamefont {C.}~\bibnamefont {Mora}}, \ and\ \bibinfo
  {author} {\bibfnamefont {R.}~\bibnamefont {Egger}},\ }\href@noop {}
  {\bibfield  {journal} {\bibinfo  {journal} {New Journal of Physics}\ }\textbf
  {\bibinfo {volume} {7}},\ \bibinfo {pages} {192} (\bibinfo {year}
  {2005})}\BibitemShut {NoStop}%
\bibitem [{\citenamefont {Melezhik}\ and\ \citenamefont
  {Schmelcher}(2009)}]{melezhik2009}%
  \BibitemOpen
  \bibfield  {author} {\bibinfo {author} {\bibfnamefont {V.~S.}\ \bibnamefont
  {Melezhik}}\ and\ \bibinfo {author} {\bibfnamefont {P.}~\bibnamefont
  {Schmelcher}},\ }\href@noop {} {\bibfield  {journal} {\bibinfo  {journal}
  {New Journal of Physics}\ }\textbf {\bibinfo {volume} {11}},\ \bibinfo
  {pages} {073031} (\bibinfo {year} {2009})}\BibitemShut {NoStop}%
\bibitem [{\citenamefont {Giannakeas}\ \emph {et~al.}(2013)\citenamefont
  {Giannakeas}, \citenamefont {Melezhik},\ and\ \citenamefont
  {Schmelcher}}]{giannakeas2013}%
  \BibitemOpen
  \bibfield  {author} {\bibinfo {author} {\bibfnamefont {P.}~\bibnamefont
  {Giannakeas}}, \bibinfo {author} {\bibfnamefont {V.~S.}\ \bibnamefont
  {Melezhik}}, \ and\ \bibinfo {author} {\bibfnamefont {P.}~\bibnamefont
  {Schmelcher}},\ }\href@noop {} {\bibfield  {journal} {\bibinfo  {journal}
  {Phys. Rev. Lett.}\ }\textbf {\bibinfo {volume} {111}},\ \bibinfo {pages}
  {183201} (\bibinfo {year} {2013})}\BibitemShut {NoStop}%
\bibitem [{\citenamefont {Sinha}\ and\ \citenamefont
  {Santos}(2007)}]{sinha2007}%
  \BibitemOpen
  \bibfield  {author} {\bibinfo {author} {\bibfnamefont {S.}~\bibnamefont
  {Sinha}}\ and\ \bibinfo {author} {\bibfnamefont {L.}~\bibnamefont {Santos}},\
  }\href@noop {} {\bibfield  {journal} {\bibinfo  {journal} {Phys. Rev. Lett.}\
  }\textbf {\bibinfo {volume} {99}},\ \bibinfo {pages} {140406} (\bibinfo
  {year} {2007})}\BibitemShut {NoStop}%
\bibitem [{\citenamefont {Hanna}\ \emph {et~al.}(2012)\citenamefont {Hanna},
  \citenamefont {Tiesinga}, \citenamefont {Mitchell},\ and\ \citenamefont
  {Julienne}}]{hanna2012}%
  \BibitemOpen
  \bibfield  {author} {\bibinfo {author} {\bibfnamefont {T.~M.}\ \bibnamefont
  {Hanna}}, \bibinfo {author} {\bibfnamefont {E.}~\bibnamefont {Tiesinga}},
  \bibinfo {author} {\bibfnamefont {W.~F.}\ \bibnamefont {Mitchell}}, \ and\
  \bibinfo {author} {\bibfnamefont {P.~S.}\ \bibnamefont {Julienne}},\
  }\href@noop {} {\bibfield  {journal} {\bibinfo  {journal} {Phys. Rev. A}\
  }\textbf {\bibinfo {volume} {85}},\ \bibinfo {pages} {022703} (\bibinfo
  {year} {2012})}\BibitemShut {NoStop}%
\bibitem [{\citenamefont {Petrov}\ and\ \citenamefont
  {Shlyapnikov}(2001)}]{petrov2001}%
  \BibitemOpen
  \bibfield  {author} {\bibinfo {author} {\bibfnamefont {D.~S.}\ \bibnamefont
  {Petrov}}\ and\ \bibinfo {author} {\bibfnamefont {G.~V.}\ \bibnamefont
  {Shlyapnikov}},\ }\href@noop {} {\bibfield  {journal} {\bibinfo  {journal}
  {Phys. Rev. A}\ }\textbf {\bibinfo {volume} {64}},\ \bibinfo {pages} {012706}
  (\bibinfo {year} {2001})}\BibitemShut {NoStop}%
\bibitem [{\citenamefont {Idziaszek}\ and\ \citenamefont
  {Calarco}(2006)}]{idziaszek2006}%
  \BibitemOpen
  \bibfield  {author} {\bibinfo {author} {\bibfnamefont {Z.}~\bibnamefont
  {Idziaszek}}\ and\ \bibinfo {author} {\bibfnamefont {T.}~\bibnamefont
  {Calarco}},\ }\href@noop {} {\bibfield  {journal} {\bibinfo  {journal} {Phys.
  Rev. Lett.}\ }\textbf {\bibinfo {volume} {96}},\ \bibinfo {pages} {013201}
  (\bibinfo {year} {2006})}\BibitemShut {NoStop}%
\bibitem [{\citenamefont {Zhang}\ and\ \citenamefont
  {Greene}(2013)}]{zhang2013}%
  \BibitemOpen
  \bibfield  {author} {\bibinfo {author} {\bibfnamefont {C.}~\bibnamefont
  {Zhang}}\ and\ \bibinfo {author} {\bibfnamefont {C.~H.}\ \bibnamefont
  {Greene}},\ }\href@noop {} {\bibfield  {journal} {\bibinfo  {journal} {Phys.
  Rev. A}\ }\textbf {\bibinfo {volume} {88}},\ \bibinfo {pages} {012715}
  (\bibinfo {year} {2013})}\BibitemShut {NoStop}%
\bibitem [{\citenamefont {Zhang}\ and\ \citenamefont {Greene}()}]{zhang2013b}%
  \BibitemOpen
  \bibfield  {author} {\bibinfo {author} {\bibfnamefont {C.}~\bibnamefont
  {Zhang}}\ and\ \bibinfo {author} {\bibfnamefont {C.~H.}\ \bibnamefont
  {Greene}},\ }\href@noop {} {}\bibinfo {note} {{a}rXiv:1312.6666,
  (2013)}\BibitemShut {NoStop}%
\bibitem [{\citenamefont {Cui}\ \emph {et~al.}(2010)\citenamefont {Cui},
  \citenamefont {Wang},\ and\ \citenamefont {Zhou}}]{cui2010}%
  \BibitemOpen
  \bibfield  {author} {\bibinfo {author} {\bibfnamefont {X.}~\bibnamefont
  {Cui}}, \bibinfo {author} {\bibfnamefont {Y.}~\bibnamefont {Wang}}, \ and\
  \bibinfo {author} {\bibfnamefont {F.}~\bibnamefont {Zhou}},\ }\href@noop {}
  {\bibfield  {journal} {\bibinfo  {journal} {Phys. Rev. Lett.}\ }\textbf
  {\bibinfo {volume} {104}},\ \bibinfo {pages} {153201} (\bibinfo {year}
  {2010})}\BibitemShut {NoStop}%
\bibitem [{\citenamefont {Fedichev}\ \emph {et~al.}(2004)\citenamefont
  {Fedichev}, \citenamefont {Bijlsma},\ and\ \citenamefont
  {Zoller}}]{fedichev2004}%
  \BibitemOpen
  \bibfield  {author} {\bibinfo {author} {\bibfnamefont {P.~O.}\ \bibnamefont
  {Fedichev}}, \bibinfo {author} {\bibfnamefont {M.~J.}\ \bibnamefont
  {Bijlsma}}, \ and\ \bibinfo {author} {\bibfnamefont {P.}~\bibnamefont
  {Zoller}},\ }\href@noop {} {\bibfield  {journal} {\bibinfo  {journal} {Phys.
  Rev. Lett.}\ }\textbf {\bibinfo {volume} {92}},\ \bibinfo {pages} {080401}
  (\bibinfo {year} {2004})}\BibitemShut {NoStop}%
\bibitem [{\citenamefont {Nishida}\ and\ \citenamefont
  {Tan}(2010)}]{nishida2010}%
  \BibitemOpen
  \bibfield  {author} {\bibinfo {author} {\bibfnamefont {Y.}~\bibnamefont
  {Nishida}}\ and\ \bibinfo {author} {\bibfnamefont {S.}~\bibnamefont {Tan}},\
  }\href@noop {} {\bibfield  {journal} {\bibinfo  {journal} {Phys. Rev. A}\
  }\textbf {\bibinfo {volume} {82}},\ \bibinfo {pages} {062713} (\bibinfo
  {year} {2010})}\BibitemShut {NoStop}%
\bibitem [{\citenamefont {Kim}\ \emph {et~al.}(2005)\citenamefont {Kim},
  \citenamefont {Schmiedmayer},\ and\ \citenamefont {Schmelcher}}]{kim2005}%
  \BibitemOpen
  \bibfield  {author} {\bibinfo {author} {\bibfnamefont {J.~I.}\ \bibnamefont
  {Kim}}, \bibinfo {author} {\bibfnamefont {J.}~\bibnamefont {Schmiedmayer}}, \
  and\ \bibinfo {author} {\bibfnamefont {P.}~\bibnamefont {Schmelcher}},\
  }\href@noop {} {\bibfield  {journal} {\bibinfo  {journal} {Phys. Rev. A}\
  }\textbf {\bibinfo {volume} {72}},\ \bibinfo {pages} {042711} (\bibinfo
  {year} {2005})}\BibitemShut {NoStop}%
\bibitem [{\citenamefont {Gao}(2009)}]{gao2009}%
  \BibitemOpen
  \bibfield  {author} {\bibinfo {author} {\bibfnamefont {B.}~\bibnamefont
  {Gao}},\ }\href {\doibase 10.1103/PhysRevA.80.012702} {\bibfield  {journal}
  {\bibinfo  {journal} {Phys. Rev. A}\ }\textbf {\bibinfo {volume} {80}},\
  \bibinfo {pages} {012702} (\bibinfo {year} {2009})}\BibitemShut {NoStop}%
\bibitem [{\citenamefont {Harmin}(1982{\natexlab{a}})}]{harmin1982}%
  \BibitemOpen
  \bibfield  {author} {\bibinfo {author} {\bibfnamefont {D.~A.}\ \bibnamefont
  {Harmin}},\ }\href {\doibase 10.1103/PhysRevA.26.2656} {\bibfield  {journal}
  {\bibinfo  {journal} {Phys. Rev. A}\ }\textbf {\bibinfo {volume} {26}},\
  \bibinfo {pages} {2656} (\bibinfo {year} {1982}{\natexlab{a}})}\BibitemShut
  {NoStop}%
\bibitem [{\citenamefont {Harmin}(1982{\natexlab{b}})}]{harmin1982prl}%
  \BibitemOpen
  \bibfield  {author} {\bibinfo {author} {\bibfnamefont {D.~A.}\ \bibnamefont
  {Harmin}},\ }\href {\doibase 10.1103/PhysRevLett.49.128} {\bibfield
  {journal} {\bibinfo  {journal} {Phys. Rev. Lett.}\ }\textbf {\bibinfo
  {volume} {49}},\ \bibinfo {pages} {128} (\bibinfo {year}
  {1982}{\natexlab{b}})}\BibitemShut {NoStop}%
\bibitem [{\citenamefont {Harmin}(1985)}]{harmin1985electric}%
  \BibitemOpen
  \bibfield  {author} {\bibinfo {author} {\bibfnamefont {D.~A.}\ \bibnamefont
  {Harmin}},\ }\href@noop {} {\bibfield  {journal} {\bibinfo  {journal}
  {Comments At. Mol. Phys}\ }\textbf {\bibinfo {volume} {15}},\ \bibinfo
  {pages} {281} (\bibinfo {year} {1985})}\BibitemShut {NoStop}%
\bibitem [{\citenamefont {Greene}(1987)}]{greene1987}%
  \BibitemOpen
  \bibfield  {author} {\bibinfo {author} {\bibfnamefont {C.~H.}\ \bibnamefont
  {Greene}},\ }\href {\doibase 10.1103/PhysRevA.36.4236} {\bibfield  {journal}
  {\bibinfo  {journal} {Phys. Rev. A}\ }\textbf {\bibinfo {volume} {36}},\
  \bibinfo {pages} {4236} (\bibinfo {year} {1987})}\BibitemShut {NoStop}%
\bibitem [{\citenamefont {Aymar}\ \emph {et~al.}(1996)\citenamefont {Aymar},
  \citenamefont {Greene},\ and\ \citenamefont
  {Luc-Koenig}}]{aymar1996multichannel}%
  \BibitemOpen
  \bibfield  {author} {\bibinfo {author} {\bibfnamefont {M.}~\bibnamefont
  {Aymar}}, \bibinfo {author} {\bibfnamefont {C.~H.}\ \bibnamefont {Greene}}, \
  and\ \bibinfo {author} {\bibfnamefont {E.}~\bibnamefont {Luc-Koenig}},\
  }\href@noop {} {\bibfield  {journal} {\bibinfo  {journal} {Rev. Mod. Phys.}\
  }\textbf {\bibinfo {volume} {68}},\ \bibinfo {pages} {1015} (\bibinfo {year}
  {1996})}\BibitemShut {NoStop}%
\bibitem [{\citenamefont {Lupu-Sax}(1998)}]{lupu1998quantum}%
  \BibitemOpen
  \bibfield  {author} {\bibinfo {author} {\bibfnamefont {A.}~\bibnamefont
  {Lupu-Sax}},\ }\emph {\bibinfo {title} {Quantum scattering theory and
  applications}},\ \href@noop {} {Ph.D. thesis},\ \bibinfo  {school} {Citeseer}
  (\bibinfo {year} {1998})\BibitemShut {NoStop}%
\bibitem [{\citenamefont {Landau}\ and\ \citenamefont
  {Lifshits}(1977)}]{landau1977quantum}%
  \BibitemOpen
  \bibfield  {author} {\bibinfo {author} {\bibfnamefont {L.~D.}\ \bibnamefont
  {Landau}}\ and\ \bibinfo {author} {\bibfnamefont {E.~M.}\ \bibnamefont
  {Lifshits}},\ }\href {http://books.google.de/books?id=J9ui6KwC4mMC} {\emph
  {\bibinfo {title} {Quantum Mechanics: Non-relativistic Theory}}}\ (\bibinfo
  {publisher} {Butterworth-Heinemann},\ \bibinfo {year} {1977})\BibitemShut
  {NoStop}%
\bibitem [{\citenamefont {He{\ss}}\ \emph {et~al.}()\citenamefont {He{\ss}},
  \citenamefont {Giannakeas},\ and\ \citenamefont {Schmelcher}}]{hess2014prep}%
  \BibitemOpen
  \bibfield  {author} {\bibinfo {author} {\bibfnamefont {B.}~\bibnamefont
  {He{\ss}}}, \bibinfo {author} {\bibfnamefont {P.}~\bibnamefont {Giannakeas}},
  \ and\ \bibinfo {author} {\bibfnamefont {P.}~\bibnamefont {Schmelcher}},\
  }\href@noop {} {}\bibinfo {note} {{i}n {p}reparation}\BibitemShut {NoStop}%
\bibitem [{\citenamefont {Gao}(2000)}]{gao2000}%
  \BibitemOpen
  \bibfield  {author} {\bibinfo {author} {\bibfnamefont {B.}~\bibnamefont
  {Gao}},\ }\href {\doibase 10.1103/PhysRevA.62.050702} {\bibfield  {journal}
  {\bibinfo  {journal} {Phys. Rev. A}\ }\textbf {\bibinfo {volume} {62}},\
  \bibinfo {pages} {050702} (\bibinfo {year} {2000})}\BibitemShut {NoStop}%
\bibitem [{\citenamefont {Gao}(2011)}]{gao2011}%
  \BibitemOpen
  \bibfield  {author} {\bibinfo {author} {\bibfnamefont {B.}~\bibnamefont
  {Gao}},\ }\href@noop {} {\bibfield  {journal} {\bibinfo  {journal} {Phys.
  Rev. A}\ }\textbf {\bibinfo {volume} {84}},\ \bibinfo {pages} {022706}
  (\bibinfo {year} {2011})}\BibitemShut {NoStop}%
\bibitem [{\citenamefont {Melezhik}(2012)}]{melezhik2012}%
  \BibitemOpen
  \bibfield  {author} {\bibinfo {author} {\bibfnamefont {V.~S.}\ \bibnamefont
  {Melezhik}},\ }\href@noop {} {\emph {\bibinfo {title} {Multi-Channel
  Computations in Low-Dimensional Few-Body Phyiscs}}}\ (\bibinfo  {publisher}
  {Springer-Verlag, Berlin, Heidelberg},\ \bibinfo {year} {2012})\BibitemShut
  {NoStop}%
\bibitem [{\citenamefont {Fano}(1961)}]{fano1961qparameter}%
  \BibitemOpen
  \bibfield  {author} {\bibinfo {author} {\bibfnamefont {U.}~\bibnamefont
  {Fano}},\ }\href {\doibase 10.1103/PhysRev.124.1866} {\bibfield  {journal}
  {\bibinfo  {journal} {Phys. Rev.}\ }\textbf {\bibinfo {volume} {124}},\
  \bibinfo {pages} {1866} (\bibinfo {year} {1961})}\BibitemShut {NoStop}%
\bibitem [{\citenamefont {Gao}(1998)}]{gao1998}%
  \BibitemOpen
  \bibfield  {author} {\bibinfo {author} {\bibfnamefont {B.}~\bibnamefont
  {Gao}},\ }\href {\doibase 10.1103/PhysRevA.58.4222} {\bibfield  {journal}
  {\bibinfo  {journal} {Phys. Rev. A}\ }\textbf {\bibinfo {volume} {58}},\
  \bibinfo {pages} {4222} (\bibinfo {year} {1998})}\BibitemShut {NoStop}%
\bibitem [{\citenamefont {Chin}\ \emph {et~al.}(2004)\citenamefont {Chin},
  \citenamefont {Vuletic}, \citenamefont {Kerman}, \citenamefont {Chu},
  \citenamefont {Tiesinga}, \citenamefont {Leo},\ and\ \citenamefont
  {Williams}}]{chin2004}%
  \BibitemOpen
  \bibfield  {author} {\bibinfo {author} {\bibfnamefont {C.}~\bibnamefont
  {Chin}}, \bibinfo {author} {\bibfnamefont {V.}~\bibnamefont {Vuletic}},
  \bibinfo {author} {\bibfnamefont {A.~J.}\ \bibnamefont {Kerman}}, \bibinfo
  {author} {\bibfnamefont {S.}~\bibnamefont {Chu}}, \bibinfo {author}
  {\bibfnamefont {E.}~\bibnamefont {Tiesinga}}, \bibinfo {author}
  {\bibfnamefont {P.~J.}\ \bibnamefont {Leo}}, \ and\ \bibinfo {author}
  {\bibfnamefont {C.~J.}\ \bibnamefont {Williams}},\ }\href@noop {} {\bibfield
  {journal} {\bibinfo  {journal} {Phys. Rev. A}\ }\textbf {\bibinfo {volume}
  {70}},\ \bibinfo {pages} {032701} (\bibinfo {year} {2004})}\BibitemShut
  {NoStop}%
\end{thebibliography}%

\end{document}